\documentclass[12pt]{article}

\pdfoutput=1
\usepackage[latin9]{inputenc}
\usepackage[top=80pt,bottom=85pt,left=67pt,right=67pt]{geometry}
\usepackage{amssymb,amsmath}
\usepackage{xypic,subfigure}
\usepackage{graphicx,color}
\usepackage{float}
\usepackage{cite}
\usepackage[debug,pageanchor=false]{hyperref}
\definecolor{link}{rgb}{.8,.15,.1}
\hypersetup{colorlinks=true,linkcolor=link,citecolor=link,urlcolor=link,linktocpage}

\makeatletter
\@addtoreset{equation}{section}
\makeatother

\renewcommand{\theequation}{\thesection.\arabic{equation}}

\newcommand{\beq}{\begin{equation}}
\newcommand{\eeq}{\end{equation}}
\newcommand{\bea}{\begin{eqnarray}}
\newcommand{\eea}{\end{eqnarray}}

\newcommand{\eq}{\begin{equation}}
\newcommand{\feq}{\end{equation}}
\newcommand{\eqn}{\begin{eqnarray}}
\newcommand{\feqn}{\end{eqnarray}}

\newcommand{\ma}[1]{\mbox{$\mathcal{#1}$}}

\newcommand{\mrm}[1]{\mbox{$\mathrm{#1}$}}

\begin{document}
\begin{titlepage}

\begin{center}

\vskip .5in 
\noindent

{\Large \bf{New $\ma N=(0,4)$ AdS$_3$ near-horizons in Type IIB}}

\bigskip\medskip

Federico Faedo$^a$\footnote{federico.faedo@unimi.it}, Yolanda Lozano$^b$\footnote{ylozano@uniovi.es},  Nicol\`o Petri$^b$\footnote{petrinicolo@uniovi.es} \\
\bigskip\medskip
{\small 

a: Dipartimento di Fisica, Universit\`a di Milano, and \\
INFN, Sezione di Milano, \\
Via Celoria 16, I-20133 Milano, Italy.}

\bigskip\medskip
{\small 

b: Department of Physics, University of Oviedo, \\
Avda. Federico Garcia Lorca s/n, 33007 Oviedo, Spain.}

\vskip 1.5cm 

     	{\bf Abstract }
     	\end{center}
     	\noindent

We construct new families of $\ma N=(0,4)$ $\mathrm{AdS}_3\times S^2 \times \tilde S^2 \times S^1$ backgrounds fibred over a 2d Riemann surface in Type IIB string theory. These solutions are obtained by extracting the near-horizon limit of D3-brane box configurations, consisting on D3-D5-NS5 branes ending on D5'-NS5' background branes. We relate our families of solutions to previous $\mathrm{AdS}_3\times S^2 \times \text{CY}_2$ and $\mathrm{AdS}_3\times S^3 \times S^2$ solutions to Type IIA recently constructed in the literature.  
We construct explicit 2d quiver CFTs associated to the D3-brane box configurations, and check that the central charges match the holographic result. We extend our solutions to include D7-branes, and show that a subclass of these solutions can be interpreted in terms of D3-D5-NS5 defect branes embedded in a 5d fixed point theory. This is explicitly realised by linking our solutions to a 6d domain wall that asymptotes locally the $\mathrm{AdS}_6\times S^2 \times \Sigma_2$ solution T-dual to the Brandhuber-Oz vacuum.
     	
\noindent

\vfill
\eject

\end{titlepage}

\setcounter{footnote}{0}

\tableofcontents

\setcounter{footnote}{0}
\renewcommand{\theequation}{{\rm\thesection.\arabic{equation}}}

\section{Introduction}

AdS$_3$ solutions in string theory are of key importance in the study of black holes, since they provide full-fledged backgrounds where one can make use of gauge/gravity duality to learn about their microscopical properties. In this spirit, recent progress has focused on the construction of new AdS$_3$ solutions with $(0,4)$ Poincar\'e supercharges \cite{Couzens:2017way,Macpherson:2018mif,Lozano:2019emq,Lozano:2020bxo,Faedo:2020nol}, as well as on the identification of their 2d dual CFTs \cite{Lozano:2019jza,Lozano:2019zvg,Lozano:2019ywa,Lozano:2020bxo,Faedo:2020nol}. These CFTs play a prominent role in the microscopic description of 5d black holes with AdS$_3\times S^2$ near-horizon geometries \cite{Maldacena:1997de,Vafa:1997gr,Minasian:1999qn,Castro:2008ne,Haghighat:2015ega,Couzens:2019wls,Couzens:2020aat}. They feature as well in the description of 6d $(1,0)$ CFTs deformed away from the conformal point \cite{Haghighat:2013tka,Gadde:2015tra}. 

Three-dimensional AdS backgrounds have also played an important role as holographic duals of defect CFTs. Many examples of defect CFTs have been discussed in the string theory literature (for a non-exhaustive list of references on this topic see \cite{Karch:2000gx,DeWolfe:2001pq,Bachas:2001vj,Erdmenger:2002ex,Clark:2004sb,DHoker:2007zhm,Lunin:2007ab,Gaiotto:2008sd,Jensen:2013lxa,Dibitetto:2017tve,Dibitetto:2017klx,Dibitetto:2018iar,Chen:2019qib,Chen:2020mtv}). The typical realisation is as brane intersections ending on bound states which are known to be described by AdS vacua in the near-horizon limit. The intersection breaks some of the isometries of the vacuum, producing a lower-dimensional AdS solution described by a non-trivial warping between AdS and the internal manifold. The defect describes then the boundary conditions associated to the intersection between the defect branes and the original bound state. Explicit realisations dual to AdS$_3$ solutions with $(0,4)$ supercharges feature D2-D6-NS5 defect branes ending on D4-D8 bound states \cite{Dibitetto:2018iar,Faedo:2020nol}, or D2-D4 defect branes ending on D6-NS5 bound states \cite{Dibitetto:2017tve,Dibitetto:2017klx,Faedo:2020nol}. The first examples of near-horizon geometries of this kind of intersections have been studied firstly in \cite{Boonstra:1998yu}.
In our situation, in the first aforementioned case the ambient CFT is the 5d Sp(N) fixed point theory of \cite{Seiberg:1996bd,Brandhuber:1999np}, while in the second case it is the 6d $(1,0)$ CFT dual to the AdS$_7$ solution to massless Type IIA supergravity \cite{Cvetic:2000cj}. In this paper we will discuss further realisations in Type IIB string theory.

In \cite{Lozano:2019emq} general classes of AdS$_3\times S^2\times M_4\times I$ solutions to massive Type IIA supergravity were constructed, with $M_4$ either a $\text{CY}_2$ or a K\"ahler manifold. The solutions with $M_4=\text{CY}_2$ were shown to arise in the near-horizon limit of D2-D4-D6-NS5-D8 brane intersections \cite{Faedo:2020nol}. The dual CFTs of this class of solutions were studied in \cite{Lozano:2019jza,Lozano:2019zvg,Lozano:2019ywa}. They arise in the infrared limit of $(0,4)$ quiver gauge theories containing two families of unitary gauge groups, $\Pi_{i=1}^{n} \text{U}(k_i)\times \text{U}({\tilde k}_i)$. The gauge group $\text{U}(k_i)$ is associated to $k_i$ D2-branes while the gauge group $\text{U}({\tilde k}_i)$ is associated to ${\tilde k}_i$ D6-branes, wrapped on the $\text{CY}_2$. Both D2 and D6 branes are stretched between NS5-branes. On top of this, there are D4 and D8 branes that provide flavour groups to both types of gauge groups, and render the field theory anomaly-free. A subclass of these solutions was given a surface defect interpretation in \cite{Faedo:2020nol} in terms of D2-D6-NS5 defect branes within the D4-D8 background brane system of Brandhuber-Oz \cite{Brandhuber:1999np}.

In the massless case the previous solutions can be uplifted to M-theory to produce a new class of AdS$_3\times S^3/\mathbb{Z}_k\times M_4\times I$ solutions with $\mathcal{N}=(0,4)$ supersymmetries \cite{Lozano:2020bxo}. These solutions provide explicit holographic duals to the 2d $(0,4)$ quiver gauge theories with unitary gauge groups supported by $M_A$-strings \cite{Haghighat:2013gba}. These are the strings that arise on the boundaries of M2-branes suspended between parallel M5-branes. Indeed, a subclass of these solutions admits a defect interpretation in terms of M2-M5 branes ending on an intersection of M5' branes and KK-monopoles. The new class of solutions gives rise, upon reduction to Type IIA, to a novel class of AdS$_3\times S^3/\mathbb{Z}_k\times {\tilde S}^2$ solutions fibred over two intervals with $\mathcal{N}=(0,4)$ supersymmetries \cite{Faedo:2020nol}. It was further shown in that reference that a subclass of these solutions allows for a defect interpretation in terms of D2-D4 branes within a D6-KK monopole intersection, where a 6d $(1,0)$ CFT lives. Explicit quiver CFTs realising the embedding of the D2-D4 branes in the 6d quiver CFT dual to this solution were also constructed.

In this paper we further extend the aforementioned line of research by focussing on AdS$_3$ solutions  with $\mathcal{N}=(0,4)$ supersymmetries in Type IIB string theory and their brane origin. We start in section~\ref{AdS3solutions} constructing supergravity solutions associated to D3-NS5-D5 branes intersecting with D5'-NS5' branes, that we generalise acting with an $\text{SL}(2,\mathbb{R})$ transformation. These brane configurations realise D3-brane boxes, like the ones discussed in \cite{Hanany:2018hlz} (see also \cite{Hanany:1997tb,Hanany:1998it}). The AdS$_3$ backgrounds are obtained by extracting the near-horizon limit of the aforementioned brane solutions. We then study in detail the two classes of solutions that are S-dual to each other, in the near-horizon limit. In section~\ref{TST} we relate these solutions to the AdS$_3\times S^2\times T^4\times I$ solutions to massive Type IIA supergravity constructed in \cite{Lozano:2019emq}, particularised to the massless case, and with the AdS$_3\times S^3/\mathbb{Z}_k\times {\tilde S}^2\times I\times I'$  solutions to massless Type IIA constructed in \cite{Faedo:2020nol}. By doing this we provide a unified M-theory picture in terms of the AdS$_3\times S^3/\mathbb{Z}_k \times T^4/\mathbb{Z}_{k'}\times I$ solutions constructed in \cite{Lozano:2020bxo}. In section~\ref{FT-interpretation} we discuss the 2d CFTs dual to our solutions. We construct explicit 2d quiver CFTs from the associated D3-brane box configuration. The solutions T-dual to the AdS$_3\times S^2\times T^4\times I$ solutions in \cite{Lozano:2019emq} are holographically dual to the same 2d CFTs. Instead, we obtain the 2d quiver gauge theories dual to a particular global completion of the latter through a 90 degrees rotation of the previous quivers, exchanging D5 and NS5 branes. 

In section~\ref{AdS6-defects} we set to provide a surface defect CFT interpretation to the new classes of solutions discussed in our paper. We first generalise the construction in section~\ref{D3NS5D5setup} to include D7-branes. We then show that these more general solutions admit a defect interpretation within the 5d fixed point theory associated to the T-dual of the Brandhuber-Oz solution \cite{Bergman:2012kr}.
This interpretation can be given through a map that relates our $\mrm{AdS}_3$ near-horizon to the Type IIB uplift of the domain wall solution to 6d $(1,1)$ minimal supergravity found in \cite{Dibitetto:2018iar}. This domain wall is featured by an $\mrm{AdS}_3$ slicing and it crucially flows asymptotically to an AdS$_6$ solution whose uplift to Type IIB (using the formulas in \cite{Hong:2018amk}\footnote{See also \cite{Jeong:2013jfc}.}) gives rise to the T-dual of the Brandhuber-Oz solution. Section~\ref{conclusions} contains our conclusions. Appendix~\ref{summary2dCFT} contains a summary of the AdS$_3\times S^2\times \text{CY}_2\times I$ solutions and their 2d dual CFTs following  \cite{Lozano:2019emq,Lozano:2019jza,Lozano:2019zvg,Lozano:2019ywa}, base of our field theory analysis in section~\ref{FT-interpretation}. Appendix~\ref{6dsugra} contains a summary of the main properties of the uplift to Type IIB supergravity \cite{Hong:2018amk} of the 6d domain wall solution to 6d $(1,1)$ minimal supergravity, that we use to provide the defect CFT interpretation in section~\ref{AdS6-defects}.

\section{$\ma N=(0,4)$ AdS$_3$ backgrounds in Type IIB}\label{AdS3solutions}

In this section we present a new class of $\mrm{AdS}_3$ backgrounds of Type IIB string theory with $\ma N=(0,4)$ supersymmetry. These solutions are defined by $\mathrm{AdS}_3\times S^2 \times \tilde S^2 \times S^1$ geometries  fibred over a 2d Riemann surface, and are obtained in the near-horizon limit of D3-D5-NS5 branes ending on D5'-NS5' background branes, realising a D3-brane box configuration \cite{Hanany:2018hlz}. The emergence of $\mathrm{AdS}_3$ is crucially related to the requirement that the D3-D5-NS5 brane system is fully localised in the worldvolume of the D5'-NS5' branes.
This particular distribution of D3-D5-NS5 branes breaks the $\mathrm{SO}(3)\times \mathrm{SO}(3)$ symmetry of the set-up, leaving only small $\ma N=(0,4)$ supersymmetry. We extend the previous class of solutions by acting with an $\mathrm{SL}(2,\mathbb{R})$ transformation depending on a rotation parameter $\xi$ related to the quantised charges of $(p,q)$ 5-branes.

\subsection{AdS$_3$ from D3-NS5-D5-NS5'-D5' brane intersections}\label{D3NS5D5setup}

We start considering the supergravity regime of a D3-NS5-D5 bound state ending on orthogonal NS5'-D5' branes. 
As we mentioned above, we choose D3-NS5-D5 branes completely localised inside the worldvolume of the NS5'-D5' branes. This particular charge distribution breaks the $\mathrm{SO(3)}\times \mathrm{SO(3)}$ symmetry group preserved on the worldvolume of the orthogonal 5-branes and implies that the intersection preserves only 4 real supercharges. 
We consider the following metric
\begin{equation}
\label{brane_metric_D3D5NS5D5NS5}
\begin{split}
ds_{10}^2 &= H_{\mathrm{D}5'}^{-1/2} \left[ H_{\mathrm{D}5}^{-1/2} H_{\mathrm{D}3}^{-1/2} \, ds^2_{\mathbb{R}^{1,1}} + H_{\mathrm{D}5}^{1/2} H_{\mathrm{D}3}^{1/2} H_{\mathrm{NS5}5} \bigl(dr^2 + r^2 ds^2_{S^2}\bigr) \right] \\
& + H_{\mathrm{D}5'}^{1/2} H_{\mathrm{NS}5} H_{\mathrm{D}5}^{-1/2} H_{\mathrm{D}3}^{-1/2}\, dz^2 + H_{\mathrm{D}5'}^{-1/2} H_{\mathrm{NS}5'} H_{\mathrm{D}5}^{1/2} H_{\mathrm{D}3}^{-1/2} \, d\phi^2 \\
& + H_{\mathrm{D}5'}^{1/2} H_{\mathrm{NS}5'} H_{\mathrm{D}5}^{-1/2} H_{\mathrm{D}3}^{1/2} \bigl(d\rho^2 + \rho^2 ds^2_{\tilde S^2}\bigr) \,.
\end{split}
\end{equation}
The requirement that the D3-NS5-D5 branes are fully localised in the worldvolume of the NS5'-D5' branes implies that the corresponding warp factors must be smeared, i.e.\ 
$H_{\mathrm{D}5}=H_{\mathrm{D}5}(r)$, $H_{\mathrm{NS}5}=H_{\mathrm{NS}5}(r)$ and $H_{\mathrm{D}3}=H_{\mathrm{D}3}(r)$. Moreover, we require also the smearing of the NS5' charge distribution along $\phi$. This assumption links this intersection, depicted in Table \ref{Table:branesIIB}, to the Type IIA brane set-up studied in \cite{Faedo:2020nol}, describing D2-NS5-D6 branes ending on an intersection of D4-branes with KK-monopoles. Indeed, one can check that both brane set-ups are related by T-duality along the $\phi$ direction.
\begin{table}[http!]
\renewcommand{\arraystretch}{1}
\begin{center}
\scalebox{1}[1]{
\begin{tabular}{c||c c|c c c || c | c c c c}
 branes & $t$ & $x^1$ & $r$ & $\theta^{1}$ & $\theta^{2}$ & $z$ & $\rho$ & $\varphi^1$ & $\varphi^2$ & $\phi$ \\
\hline \hline
$\mrm{D}3$ & $\times$ & $\times$ & $-$ & $-$ & $-$ & $\times$ & $-$ & $-$ & $-$ & $\times$ \\
$\mrm{D}5$ & $\times$ & $\times$ & $-$ & $-$ & $-$ & $\times$ & $\times$ & $\times$ & $\times$ & $-$ \\
$\mrm{NS}5$ & $\times$ & $\times$ & $-$ & $-$ & $-$ & $-$ & $\times$ & $\times$ & $\times$ & $\times$ \\
$\mrm{D}5'$ & $\times$ & $\times$ & $\times$ & $\times$ & $\times$ & $-$ & $-$ & $-$ & $-$ & $\times$ \\
$\mrm{NS}5'$ & $\times$ & $\times$ & $\times$  & $\times$ & $\times$ & $\times$ & $-$ & $-$ &$-$& $-$ \\
\end{tabular}
}
\end{center}
\caption{Brane picture underlying our D3-D5-NS5-D5'-NS5' brane intersection. The system describes a $\mrm{BPS}/8$ D3-brane box configuration.} \label{Table:branesIIB}
\end{table}
The gauge potentials and the dilaton associated to this particular charge distribution are given by
\begin{equation}
\begin{aligned}\label{brane_potentials_D3NS5D5NS5D5}
B_{(6)} &= H_{\mathrm{NS}5} H_{\mathrm{D}5} H_{\mathrm{NS}5'}^{-1}\,r^2\,\text{vol}_{\mathbb{R}^{1,1}} \wedge dr \wedge \text{vol}_{ S^2} \wedge dz + H_{\mathrm{NS}5'} H_{\mathrm{D}5'} H_{\mathrm{NS}5}^{-1}\,\rho^2\,\text{vol}_{\mathbb{R}^{1,1}} \wedge d\rho \wedge \text{vol}_{\tilde S^2} \wedge d\phi\,,\\ \vspace{0.4cm}
C_{(6)} &= H_{\mathrm{NS}5} H_{\mathrm{D}5} H_{\mathrm{D}5'}^{-1}\,r^2\,\text{vol}_{\mathbb{R}^{1,1}} \wedge dr \wedge \text{vol}_{ S^2}\wedge d\phi+H_{\mathrm{NS}5'} H_{\mathrm{D}5'} H_{\mathrm{D}5}^{-1}\,\rho^2\,\text{vol}_{\mathbb{R}^{1,1}} \wedge d\rho \wedge \text{vol}_{\tilde S^2} \wedge dz \,,\\
C_{(4)} &= H_{\mathrm{D}3}^{-1}\,\text{vol}_{\mathbb{R}^{1,1}} \wedge dz \wedge d\phi\,,\\
e^{\Phi} &= H_{\mathrm{D}5'}^{-1/2} H_{\mathrm{D}5}^{-1/2} H_{\mathrm{NS}5'}^{1/2} H_{\mathrm{NS}5}^{1/2}\,,
\end{aligned}
\end{equation}
where the D5' branes are taken completely localised in their transverse space, parametrised by the coordinates $(z,\rho)$. From the gauge potentials we can obtain the fluxes
\begin{equation}
\begin{aligned}\label{fluxes_D3NS5D5NS5D5}
H_{(3)} &= \partial_rH_{\mathrm{NS}5}\,r^2 \, \text{vol}_{S^2} \wedge dz + \partial_\rho H_{\mathrm{NS}5'}\,\rho^2 \, \text{vol}_{\tilde S^2} \wedge d\phi\,,\\
F_{(3)} &= -\partial_rH_{\mathrm{D}5}\,r^2 \, \text{vol}_{S^2} \wedge d\phi - \partial_\rho H_{\mathrm{D}5'}\,\rho^2 \, \text{vol}_{\tilde S^2} \wedge dz + H_{\mathrm{D}3}H_{\mathrm{NS}5}^{-1}\,H_{\mathrm{NS}5'}\,\partial_zH_{\mathrm{D}5'}\,\rho^2 \, d\rho\wedge\text{vol}_{\tilde S^2}\,,\\
F_{(5)} &= \partial_rH_{\mathrm{D}3}^{-1}\,\text{vol}_{\mathbb{R}^{1,1}}\wedge dr\wedge dz \wedge d\phi+H_{\mathrm{D}5'}H_{\mathrm{NS}5'}\,\partial_rH_{\mathrm{D}3}\,r^2\rho^2 \, \text{vol}_{S^2}\wedge \text{vol}_{\tilde S^2}\wedge d\rho\,.
\end{aligned}
\end{equation}
With the fluxes \eqref{fluxes_D3NS5D5NS5D5} the equations of motion and the Bianchi identities of Type IIB supergravity decouple in two sets. The first set describes the dynamics of the D3-NS5-D5 branes in terms of harmonic functions on $\mathbb{R}^3_r$,
\begin{equation}
\begin{split}\label{D3D5NS5EOMS}
& \nabla^2_{\mathbb{R}^3_r} H_{\mathrm{D}3}=0 \qquad \text{with}\qquad H_{\mathrm{NS}5}=H_{\mathrm{D}3}\,,\\
& \nabla^2_{\mathbb{R}^3_r} H_{\mathrm{D}5}=0\,.
\end{split}
\end{equation}
The second set of equations is associated to the NS5'-D5' branes and is given by
\begin{equation}\label{NS5D5EOMS}
\nabla^2_{\mathbb{R}^3_\rho} H_{\mathrm{D}5'} + H_{\mathrm{NS}5'} \, \partial_z^2 H_{\mathrm{D}5'}=0  \qquad  \text{and}  \qquad  \nabla^2_{\mathbb{R}^3_\rho} H_{\mathrm{NS}5'} = 0  \,.
\end{equation}
We can easily integrate the equations \eqref{D3D5NS5EOMS} and consider the particular solution
\begin{equation}\label{D3D5NS5EOMSsol}
  H_{\mathrm{D}3}(r)=H_{\mathrm{NS}5}(r)=1+\frac{Q_{\mathrm{D}3}}{r}\,,\qquad  H_{\mathrm{D}5}(r)=1+\frac{Q_{\mathrm{D}5}}{r}\,,
 \end{equation}
where it is now explicit that the conservation of charge forces the NS5 and D3 brane charges to be equal. 

We can now extract an $\mathrm{AdS}_3\times S^2 \times \tilde S^2 \times S^1\times I_\rho \times I_z$ background by taking the limit $r\to0$. This limit defines a class of solutions to Type IIB supergravity given by\footnote{We rescaled the Minkowski coordinates as $(t,x^1)\mapsto 2\,Q_{\mathrm{D}5}^{1/2}\,Q_{\mathrm{D}3}\,(t,x^1)$ in order to construct the unitary AdS$_3$ metric. Subsequently, we also rescaled the charges $Q_{\mathrm{D}3}\mapsto Q_{\mathrm{D}3}/2$, $Q_{\mathrm{D}5}\mapsto Q_{\mathrm{D}5}/2$ and the coordinate $\rho\mapsto\rho/2$ in order to be consistent with the dual field theory description provided in section~\ref{FT-interpretation}.}
\begin{equation}
\label{brane_metric_D3NS5D5NS5D5_nh}
\begin{aligned}
ds_{10}^2 &= Q_{\text{D}3}^{3/2} Q_{\text{D}5}^{1/2} H_{\mathrm{D}5'}^{-1/2}  \Bigl(ds^2_{\text{AdS}_3} + \frac14\,ds^2_{S^2}\Bigr) + Q_{\text{D}3}^{1/2} Q_{\text{D}5}^{-1/2} H_{\mathrm{D}5'}^{1/2} \, dz^2 + Q_{\text{D}3}^{-1/2} Q_{\text{D}5}^{1/2} H_{\mathrm{D}5'}^{-1/2} H_{\mathrm{NS}5'} \, d\phi^2 \\
& + \frac14\,Q_{\text{D}3}^{1/2} Q_{\text{D}5}^{-1/2} H_{\mathrm{D}5'}^{1/2} H_{\mathrm{NS}5'} \bigl(d\rho^2 + \rho^2 ds^2_{\tilde{S}^2}\bigr) \,, \\
H_{(3)} &= -\frac{Q_{\text{D}3}}{2} \, \text{vol}_{S^2} \wedge dz + \frac12\,\partial_\rho H_{\mathrm{NS}5'} \,\rho^2 \, \text{vol}_{\tilde S^2} \wedge d\phi \,,  \qquad  e^{\Phi} = Q_{\text{D}3}^{1/2} Q_{\text{D}5}^{-1/2} H_{\mathrm{D}5'}^{-1/2} H_{\mathrm{NS}5'}^{1/2} \,, \\
F_{(3)} &= \frac{Q_{\text{D}5}}{2} \, \text{vol}_{S^2} \wedge d\phi - \frac12\,\partial_\rho H_{\mathrm{D}5'} \, \rho^2 \, \text{vol}_{\tilde S^2} \wedge dz + \frac18\,H_{\mathrm{NS}5'} \partial_z H_{\mathrm{D}5'} \, \rho^2 \, d\rho \wedge \text{vol}_{\tilde S^2} \,, \\
F_{(5)} &= 2Q_{\text{D}3} Q_{\text{D}5} \, \text{vol}_{\text{AdS}_3} \wedge dz \wedge d\phi -\frac{Q_{\text{D}3}}{16} H_{\mathrm{D}5'} H_{\mathrm{NS}5'} \rho^2 \, \text{vol}_{S^2} \wedge \text{vol}_{\tilde S^2} \wedge d\rho \,.
\end{aligned}
\end{equation}
Because of the smearing of all the branes along $\phi$ we can take it as defining a circle. As we will discuss in section~\ref{TST}, the solution is then related by T-duality to the AdS$_3\times S^2\times T^4/\mathbb{Z}_{k'}\times I_z$ class of solutions constructed in \cite{Lozano:2019emq}, restricted to the massless case. These solutions were obtained in the near-horizon limit of a bound state of D2-NS5-D6 branes ending on a D4-KK intersection \cite{Faedo:2020nol}.


\subsection{More general solutions through $\mrm{SL}(2,\mathbb R)$}\label{AdS3pq5}

In this section we construct an extension of the brane solution \eqref{brane_metric_D3D5NS5D5NS5} by acting with an $\mrm{SL}(2,\mathbb R)$ transformation. We introduce an $\mrm{SL}(2,\mathbb R)$ element parametrised by an angle $\xi \in[0,\frac{\pi}{2}]$ as follows
\begin{equation}
S=
\left(\begin{array}{cc}
\cos \xi & - \sin \xi \\
\sin\xi &   \cos\xi \\
\end{array}\right)\,.
 \end{equation}
Denoting by $F_{(n),s}$, $\Phi_{s}$ and $ds^2_{10,s}$ the background fields describing the seed solution given by equations \eqref{brane_metric_D3D5NS5D5NS5} and \eqref{fluxes_D3NS5D5NS5D5}, associated to D3-NS5-D5 branes ending on D5'-NS5' branes, the $S$  transformation acts as usual in Type IIB supergravity,
\begin{equation}
 \begin{split}
  & \tau=\frac{\cos\xi\,\tau_s-\sin\xi}{\sin\xi\,\tau_s+\cos\xi}\,, \qquad F_{(5)}=F_{(5),s}\,,\\\vspace{5mm}
&\left(\begin{array}{c}
\tilde F_{(3)} \\
H_{(3)}\\
\end{array}\right)=\left(\begin{array}{cc}
\cos \xi & - \sin \xi \\
\sin\xi &   \cos\xi \\
\end{array}\right)\left(\begin{array}{c}
F_{(3),s} \\
H_{(3),s}\\
\end{array}\right)\,,
 \end{split}
\end{equation}
where we introduced the axio-dilaton $\tau=C_{(0)}+ie^{-\Phi}$. The seed background has vanishing axion, and with this transformation a profile for $C_{(0)}$ is generated. Given this, the 3-form flux describing the rotated solution is given by $F_{(3)}=\tilde F_{(3)}-C_{(0)}H_{(3)}$.
Finally, since we are working in string frame, we have $ds^2_{10}=|\cos\xi+\sin\xi\,\tau|\,ds^2_{10,s}$. Acting with these  transfomations on equations \eqref{brane_metric_D3D5NS5D5NS5} and \eqref{fluxes_D3NS5D5NS5D5}, we obtain for the metric
\begin{equation}
\label{brane_metric_D3D5NS5D5NS5rotated}
\begin{aligned}
ds_{10}^2 &= \Delta^{1/2} \Bigl[ H_{\mathrm{D}5'}^{-1/2} \left( H_{\mathrm{D}5}^{-1/2} H_{\mathrm{D}3}^{-1/2} \, ds^2_{\mathbb{R}^{1,1}} + H_{\mathrm{D}5}^{1/2} H_{\mathrm{D}3}^{1/2} H_{\mathrm{NS5}5} \bigl(dr^2 + r^2 ds^2_{S^2}\bigr) \right) \\
& + H_{\mathrm{D}5'}^{1/2} H_{\mathrm{NS}5} H_{\mathrm{D}5}^{-1/2} H_{\mathrm{D}3}^{-1/2} \, dz^2 + H_{\mathrm{D}5'}^{-1/2} H_{\mathrm{NS}5'} H_{\mathrm{D}5}^{1/2} H_{\mathrm{D}3}^{-1/2} \, d\phi^2 \\
& + H_{\mathrm{D}5'}^{1/2} H_{\mathrm{NS}5'} H_{\mathrm{D}5}^{-1/2} H_{\mathrm{D}3}^{1/2} \bigl(d\rho^2 + \rho^2 ds^2_{\tilde S^2}\bigl) \Bigr] \,, \\
\Delta =& \ c^2 + \frac{H_{\mathrm{D}5'}}{H_{\mathrm{NS}5'}}\frac{H_{\mathrm{D}5}}{H_{\mathrm{NS}5}}\,s^2 \,,
\end{aligned}
\end{equation}
with $s=\sin\xi$ and $c=\cos\xi$. The dilaton and the new $C_{(0)}$ fields can be extracted from the axio-dilaton,
\begin{equation}
e^{\Phi}=\Delta\,H_{\mathrm{D}5'}^{-1/2}H_{\mathrm{D}5}^{-1/2}H_{\mathrm{NS}5'}^{1/2}H_{\mathrm{NS}5}^{1/2}\,, \qquad C_{(0)} = sc\,\Delta^{-1} \left(\frac{H_{\mathrm{D}5'}}{H_{\mathrm{NS}5'}}\frac{H_{\mathrm{D}5}}{H_{\mathrm{NS}5}} -1 \right)\,.
\end{equation}
From the expression above we see that the axion vanishes in the two extremes, i.e.\ $\xi=0,\frac{\pi}{2}$. Finally, the fluxes have the form
\begin{equation}
\begin{aligned}\label{fluxes_D3NS5D5NS5D5rotated}
H_{(3)} &= c\,\partial_r H_{\mathrm{NS}5}\,r^2 \, \text{vol}_{S^2} \wedge dz + c\,\partial_\rho H_{\mathrm{NS}5'}\,\rho^2 \, \text{vol}_{\tilde S^2} \wedge d\phi - s\,\partial_r H_{\mathrm{D}5}\,r^2 \, \text{vol}_{S^2} \wedge d\phi \\
& - s\,\partial_\rho H_{\mathrm{D}5'}\,\rho^2 \, \text{vol}_{\tilde S^2} \wedge dz + s\,H_{\mathrm{D}3} H_{\mathrm{NS}5}^{-1} H_{\mathrm{NS}5'} \partial_z H_{\mathrm{D}5'}\,\rho^2 \, d\rho \wedge \text{vol}_{\tilde S^2} \,,\\
F_{(1)} &= sc\,\Delta^{-2} \Bigl[ H_{\mathrm{D}5'} H_{\mathrm{NS}5'}^{-1} H_{\mathrm{NS}5}^{-1} \left( \partial_r H_{\mathrm{D}5} - H_{\mathrm{D}5} H_{\mathrm{NS}5}^{-1} \partial_r H_{\mathrm{NS}5} \right) dr + H_{\mathrm{D}5} H_{\mathrm{NS}5}^{-1} H_{\mathrm{NS}5'}^{-1} \partial_z H_{\mathrm{D}5'} dz \\
& + H_{\mathrm{D}5} H_{\mathrm{NS}5}^{-1} H_{\mathrm{NS}5'}^{-1} \left( \partial_\rho H_{\mathrm{D}5'} - H_{\mathrm{D}5'} H_{\mathrm{NS}5'}^{-1} \partial_\rho H_{\mathrm{NS}5'} \right) d\rho \Bigr] \,,\\
F_{(3)} &= \Delta^{-1} \Bigl[ -c\,\partial_r H_{\mathrm{D}5}\,r^2 \, \text{vol}_{S^2} \wedge d\phi - c\,\partial_\rho H_{\mathrm{D}5'}\,\rho^2 \, \text{vol}_{\tilde S^2} \wedge dz \\
& + c\,H_{\mathrm{D}3} H_{\mathrm{NS}5}^{-1} H_{\mathrm{NS}5'} \partial_z H_{\mathrm{D}5'}\,\rho^2 \, d\rho \wedge \text{vol}_{\tilde S^2} - s\,\frac{H_{\mathrm{D}5'}}{H_{\mathrm{NS}5'}}\frac{H_{\mathrm{D}5}}{H_{\mathrm{NS}5}}\,\partial_r H_{\mathrm{NS}5}\,r^2 \, \text{vol}_{S^2} \wedge dz \\
& - s\,\frac{H_{\mathrm{D}5'}}{H_{\mathrm{NS}5'}} \frac{H_{\mathrm{D}5}}{H_{\mathrm{NS}5}}\,\partial_\rho H_{\mathrm{NS}5'}\,\rho^2 \, \text{vol}_{\tilde S^2} \wedge d\phi \Bigr] \,, \\
F_{(5)} &= \partial_r H_{\mathrm{D}3}^{-1} \, \text{vol}_{\mathbb{R}^{1,1}} \wedge dr \wedge dz \wedge d\phi + H_{\mathrm{D}5'} H_{\mathrm{NS}5'} \partial_r H_{\mathrm{D}3}\,r^2 \rho^2 \, \text{vol}_{S^2} \wedge \text{vol}_{\tilde S^2} \wedge d\rho \,.
\end{aligned}
\end{equation}
The equations of motion and Bianchi identities are equivalent to the following system of equations,
\begin{equation}
\begin{aligned}\label{D3D5NS5EOMSrotated}
& \nabla^2_{\mathbb{R}^3_r} H_{\mathrm{D}3} = 0,  \qquad  \text{with}  \qquad  H_{\mathrm{NS}5} = H_{\mathrm{D}3}  \qquad  \text{and}  \qquad  \nabla^2_{\mathbb{R}^3_r} H_{\mathrm{D}5} = 0 \,, \\
& \nabla^2_{\mathbb{R}^3_\rho} H_{\mathrm{D}5'} + H_{\mathrm{NS}5'} \, \partial_z^2 H_{\mathrm{D}5'} = 0  \qquad  \text{and}  \qquad  \nabla^2_{\mathbb{R}^3_\rho} H_{\mathrm{NS}5'} = 0 \,.
\end{aligned}
\end{equation}
Looking at the fluxes \eqref{fluxes_D3NS5D5NS5D5rotated} it is manifest that the $S$ transformation mixes the charge distributions of D5 and NS5 branes along the $(z,\phi)$ coordinates with the charge distributions of the NS5' and D5' branes along the $(z, \rho, \phi)$ coordinates. This implies that  5-branes charge conservation for the D5-NS5 and D5'-NS5' brane set-ups is satisfied, respectively, along a combination of the $(z,\phi)$ and $(z, \rho, \phi)$ coordinates. The string background describes then an intersection of orthogonal $(p,q)$ 5-branes with D3 branes.

If we now consider the particular solution with $H_{\mathrm{D}3}$ and $H_{\mathrm{D}5}$ given by
\begin{equation}
H_{\mathrm{D}3}(r) = 1 + \frac{Q_{\mathrm{D}3}}{r} \,,  \qquad  H_{\mathrm{D}5}(r) = 1 + \frac{Q_{\mathrm{D}5}}{r} \,,
\end{equation}
the near-horizon limit $r\to0$ reproduces an $\mathrm{AdS}_3\times S^2 \times \tilde S^2 \times S^1\times I_\rho \times I_z$ geometry, with the explicit form\footnote{We rescaled the Minkowski coordinates as $(t,x^1)\mapsto 2\,Q_{\mathrm{D}3}^{3/2}\,(t,x^1)$, along with the charges $Q_{\mathrm{D}3}\mapsto Q_{\mathrm{D}3}/2$, $Q_{\mathrm{D}5}\mapsto Q_{\mathrm{D}5}/2$ and the coordinate $\rho\mapsto\rho/2$ as we did for the seed solution in~\eqref{brane_metric_D3NS5D5NS5D5_nh}.}
\begin{equation} \label{brane_metric_D3D5NS5D5NS5rotated_nh}
\begin{aligned}
ds_{10}^2 &= \Delta^{1/2} \biggl[ Q_{\text{D}3}^{3/2} Q_{\text{D}5}^{1/2} H_{\mathrm{D}5'}^{-1/2}  \Bigl(ds^2_{\text{AdS}_3} + \frac14 ds^2_{S^2}\Bigr) + Q_{\text{D}3}^{1/2} Q_{\text{D}5}^{-1/2} H_{\mathrm{D}5'}^{1/2} dz^2 + Q_{\text{D}3}^{-1/2} Q_{\text{D}5}^{1/2} H_{\mathrm{D}5'}^{-1/2} H_{\mathrm{NS}5'} d\phi^2 \\
& + \frac14\,Q_{\text{D}3}^{1/2} Q_{\text{D}5}^{-1/2} H_{\mathrm{D}5'}^{1/2} H_{\mathrm{NS}5'} \bigl(d\rho^2 + \rho^2 ds^2_{\tilde{S}^2}\bigr) \biggr] \,,\\
e^{\Phi} &= \Delta \, Q_{\text{D}3}^{1/2} Q_{\text{D}5}^{-1/2} H_{\mathrm{D}5'}^{-1/2}H_{\mathrm{NS}5'}^{1/2}  \qquad  \text{with}  \qquad  \Delta = c^2 + \frac{Q_{\text{D}5}}{Q_{\text{D}3}} \, \frac{H_{\mathrm{D}5'}}{H_{\mathrm{NS}5'}}\,s^2 \,,\\
H_{(3)} & = -\frac12\,c\,Q_{\mathrm{D}3} \, \text{vol}_{S^2} \wedge dz + \frac12\,s\,Q_{\mathrm{D}5} \, \text{vol}_{S^2} \wedge d\phi + \frac12\,c\,\partial_\rho H_{\mathrm{NS}5'}\,\rho^2 \, \text{vol}_{\tilde S^2} \wedge d\phi \\
& - \frac12\,s\,\partial_\rho H_{\mathrm{D}5'}\,\rho^2 \, \text{vol}_{\tilde S^2} \wedge dz + \frac18\,s\,H_{\mathrm{NS}5'} \partial_z H_{\mathrm{D}5'}\,\rho^2 \, d\rho \wedge \text{vol}_{\tilde S^2} \,, \\
F_{(1)} &= sc\,\Delta^{-2} \frac{Q_{\text{D}5}}{Q_{\text{D}3}} \Bigl[ H_{\mathrm{NS}5'}^{-1} \partial_z H_{\mathrm{D}5'} \, dz + H_{\mathrm{NS}5'}^{-1} \left( \partial_\rho H_{\mathrm{D}5'} - H_{\mathrm{D}5'} H_{\mathrm{NS}5'}^{-1} \partial_\rho H_{\mathrm{NS}5'} \right) d\rho \Bigr] \,, \\
F_{(3)} &= \Delta^{-1} \Bigl[ \frac12\,c\,Q_{\mathrm{D}5} \, \text{vol}_{S^2} \wedge d\phi +
\frac12\,s\,Q_{\mathrm{D}5} H_{\mathrm{D}5'} H_{\mathrm{NS}5'}^{-1} \, \text{vol}_{S^2} \wedge dz - \frac12\,c\,\partial_\rho H_{\mathrm{D}5'}\,\rho^2 \, \text{vol}_{\tilde S^2} \wedge dz \\
& + \frac18\,c\,H_{\mathrm{NS}5'} \partial_z H_{\mathrm{D}5'}\,\rho^2 \, d\rho \wedge \text{vol}_{\tilde S^2} - \frac12\,s\,\frac{Q_{\text{D}5}}{Q_{\text{D}3}} \frac{H_{\mathrm{D}5'}}{H_{\mathrm{NS}5'}}\,\partial_\rho H_{\mathrm{NS}5'}\,\rho^2 \, \text{vol}_{\tilde S^2} \wedge d\phi \Bigr] \,, \\
F_{(5)} &= 2Q_{\text{D}3} Q_{\text{D}5} \, \text{vol}_{\text{AdS}_3} \wedge dz \wedge d\phi -\frac{Q_{\text{D}3}}{16} H_{\mathrm{D}5'} H_{\mathrm{NS}5'} \rho^2 \, \text{vol}_{S^2} \wedge \text{vol}_{\tilde S^2} \wedge d\rho \,,\\
\end{aligned}
\end{equation}
where $H_{\mathrm{D}5'}$ and $H_{\mathrm{NS}5'}$ must satisfy the conditions
\begin{equation} \label{D3D5NS5EOMSrotated_nh}
\nabla^2_{\mathbb{R}^3_\rho} H_{\mathrm{D}5'} + \frac14\,H_{\mathrm{NS}5'} \partial_z^2 H_{\mathrm{D}5'} = 0  \qquad  \text{and}  \qquad  \nabla^2_{\mathbb{R}^3_\rho} H_{\mathrm{NS}5'} = 0 \,.
\end{equation}
From now on we shall take $H_{\mathrm{NS}5'}=Q_{\mathrm{NS}5'}/\rho$. Since we rotated all the branes of Table \ref{Table:branesIIB}, this background still preserves 8 real supercharges organised in $\ma N=(0,4)$ supersymmetry. It is easy to see that if $\xi=0$ we recover the solution \eqref{brane_metric_D3NS5D5NS5D5_nh}, while for $\xi=\frac{\pi}{2}$ we get its S-dual solution. We discuss in the next section that this solution is related via T-duality to the AdS$_3\times S^3/\mathbb{Z}_k\times S^2\times I_z\times I_\rho$ class of solutions constructed in  \cite{Faedo:2020nol}, obtained in the near-horizon limit of a bound state of D2-D4 branes ending on a D6-NS5-KK brane system.

\subsection{Holographic central charge}\label{holocc}

In this subsection we derive the holographic central charge of the $\mrm{AdS}_3$ backgrounds defined by~\eqref{brane_metric_D3D5NS5D5NS5rotated_nh}. The standard prescription links the inverse of the effective Newton constant in 3d with the number of degrees of freedom of the 2d dual CFT. We extract $c_{hol}$ by following~\cite{Klebanov:2007ws,Macpherson:2014eza,Bea:2015fja}. 

Taking a generic dilaton and metric of the form,
\begin{equation}
ds_{10}^2= a(\zeta,\vec{\theta})\left(dx_{1,d}^2 + b(\zeta) \, d\zeta^2\right) + g_{ij}(\zeta,\vec{\theta}) \, d\theta^id\theta^j\,, \qquad  \Phi=\Phi(\zeta,\vec{\theta})\,,
\end{equation}
the central charge is obtained computing the auxiliary quantity 
\begin{equation}
\hat{H}= \left(\int d\vec{\theta} \sqrt{e^{-4\Phi} \det[g_{ij}] a(\zeta,\vec{\theta})^d } \right)^2,\label{Hhat}
\end{equation}
as 
\begin{equation}
c_{hol}=  \frac{3\, d^d}{ G_{\mathrm{N}}} \frac{b(\zeta)^{d/2} (\hat{H})^\frac{2d+1}{2}  }{(\hat{H}')^d} ,\label{centralx}
\end{equation}
where $G_{\mathrm{N}}$ is the Newton's constant in ten dimensions, $G_{\mathrm{N}}=8\pi^6$. For the backgrounds defined by \eqref{brane_metric_D3D5NS5D5NS5rotated_nh} we have
\begin{equation}
d=1\,,\qquad a(\zeta, \vec \theta)= \Delta^{1/2}\,Q_{\mathrm{D}3}^{3/2}\,Q_{\mathrm{D}5}^{1/2}\,H_{\mathrm{D}5'}^{-1/2}\,\zeta^{2}\,,\qquad b(\zeta)=\zeta^{-4}\,,
\end{equation}
where $\zeta$ is the radial coordinate for AdS$_3$ written in the Poincar\'e parametrisation. The final expression for the central charge is given by,
\begin{equation}\label{cholxiind}
c_{hol}=\frac{3}{2^{8}\,\pi^{5}} \,Q_{\mathrm{D}3}^{2}\,Q_{\mathrm{D}5} \text{Vol}_{S^2}\text{Vol}_{\tilde S^2} \int d\phi\,d\rho\,dz \,\rho^2H_{\mathrm{D}5'}(z,\rho)H_{\mathrm{NS}5'}(\rho)\,.
\end{equation}
From this expression it is manifest that the holographic central charge does not depend on the parameter $\xi$ and is thus S-duality invariant.

\section{Relation to $\mathcal{N}=(0,4)$ AdS$_3$ solutions in Type IIA} \label{TST}

In this section we focus on the two S-dual backgrounds arising from the class of  $\mathrm{AdS}_3\times S^2 \times \tilde S^2 \times S^1$ solutions obtained in \eqref{brane_metric_D3D5NS5D5NS5rotated_nh}, and relate them to known AdS$_3$ solutions with $\mathcal{N}=(0,4)$ supersymmetries in Type IIA through T-duality. 

The case $\xi=0$ has been discussed in section~\ref{D3NS5D5setup}. It consists on D3-D5-NS5 branes ending on D5'-NS5' background branes. The system is described by the brane solution \eqref{brane_metric_D3D5NS5D5NS5} and in the near-horizon by \eqref{brane_metric_D3NS5D5NS5D5_nh}. The latter is a $\ma N=(0,4)$ $\mathrm{AdS}_3\times S^2 \times \tilde S^2 \times S_{\phi}^1\times I_\rho \times I_z$ background defined by the functions $H_{\mathrm{D}5'}(z,\rho)$ and $H_{\mathrm{NS}5'}(\rho)$.
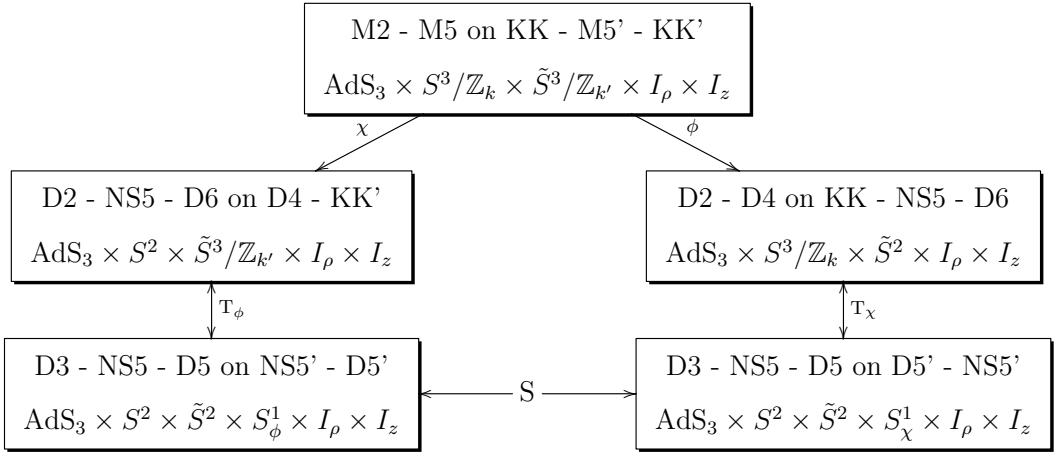
\begin{figure}[http!]
\begin{center}
\scalebox{0.9}[0.9]{ \xymatrix@C-6pc {\text{  } & *+[F-,]{\begin{array}{c} \textrm{M2 - M5  on KK  - M5' - KK'}\vspace{2mm} \\ \textrm{AdS}_3\times S^3/\mathbb{Z}_k  \times \tilde{S}^3/\mathbb{Z}_{k'} \times I_\rho \times I_z \end{array}} \ar[dl]_{\chi}\ar[dr]^{\phi} & \text{  }   \\ 
*+[F-,]{\begin{array}{c} \textrm{ D2 - NS5 - D6 on D4 - KK' } \vspace{2mm} \\ \textrm{AdS}_3\times S^2  \times \tilde S^3/\mathbb{Z}_{k'}\times I_\rho \times I_z  \end{array}}\ar@{<->}[d]^{\mathrm{T}_\phi} &  \text{  }  & *+[F-,]{\begin{array}{c} \textrm{D2 - D4 on KK - NS5 - D6} \vspace{2mm} \\ \textrm{AdS}_3\times S^3/\mathbb{Z}_k  \times \tilde{S}^2\times I_\rho \times I_z  \end{array}}\ar@{<->}[d]^{\mathrm{T}_\chi}\\
*+[F-,]{\begin{array}{c} \textrm{D3 - NS5 - D5 on NS5' - D5'} \vspace{2mm} \\  \textrm{AdS}_3\times S^2  \times \tilde S^2\times S^1_\phi\times I_\rho \times I_z \end{array}} &  \ar[l] \text{S} \ar[r] & *+[F-,]{\begin{array}{c} \textrm{ D3 - NS5 - D5 on D5' - NS5' } \vspace{2mm} \\ \textrm{AdS}_3\times S^2  \times \tilde S^2\times S^1_\chi\times I_\rho \times I_z  \end{array}}
 }}
\end{center}
\caption{Relations among the  $\ma N=(0,4)$ $\mathrm{AdS}_3$ backgrounds in Type II and M-theory discussed in this section. In the most general case the M-theory brane set-up can be reduced over the $T^2$ parametrised by the $(\chi, \phi)$ coordinates to describe an intersection of D3 branes with orthogonal $(p,q)$ 5-branes.}\label{fig:TSTchain} 
\end{figure}
It is easy to see that the brane solution \eqref{brane_metric_D3D5NS5D5NS5} is related by T-duality along the $\phi$ direction to a solution describing D2-D6-NS5 branes ending on D4-KK' background branes. Upon T-duality the ${\tilde S}^2$ and the $S^1_\phi$ give rise to a ${\tilde S^3}$ that is modded out by the number of NS5'-branes, which become KK'-monopoles. 
The latter brane set-up was put forward in \cite{Faedo:2020nol} as the brane intersection underlying the general class of $\mathrm{AdS}_3\times S^2 \times T^4/\mathbb{Z}_{k'}\times I_z$ geometries constructed in \cite{Lozano:2019emq}, restricted to the massless case. It is easy to see that indeed both near-horizon geometries are also related upon T-duality along the $\phi$ direction. 

The  M-theory realisation of the previous Type IIA solution is depicted in Figure \ref{fig:TSTchain} (left). The M-theory realisation is in terms of M2-M5 branes intersecting background M5' branes on an A-type singularity. The corresponding M2-M5-M5'-KK-KK' brane intersection was studied in \cite{Faedo:2020nol}. This intersection gives rise to the AdS$_3\times S^3/\mathbb{Z}_k  \times \tilde{S}^3/\mathbb{Z}_{k'} \times I_\rho \times I_z $ solutions to M-theory constructed in \cite{Lozano:2020bxo} in the near-horizon limit. From these, one recovers the $\mathrm{AdS}_3\times S^2 \times T^4/\mathbb{Z}_{k'}\times I_z$  solutions in  \cite{Lozano:2019emq} upon reduction along the Hopf-fibre direction of the $S^3/\mathbb{Z}_k$.


It was shown in \cite{Faedo:2020nol} that the previous M-theory backgrounds can instead be reduced along the Hopf-fibre of the second ${\tilde S}^3/\mathbb{Z}_{k'}$ sphere, to produce a new class of AdS$_3$ solutions in Type IIA that also preserve $\mathcal{N}=(0,4)$ supersymmetries. The brane set-up underlying these solutions is depicted in Figure \ref{fig:TSTchain} (right). The Type IIA vacua generated by this reduction have the form \cite{Faedo:2020nol},
\begin{equation}
\label{brane_metric_D2D4KKNS5D6_nh}
\begin{split}
ds_{10}^2 &= 4\sqrt{2} \, k \, Q_{\text{D}4} H_{\mathrm{D}6}^{-1/2} \left[ds^2_{\text{AdS}_3} + ds^2_{S^3/\mathbb{Z}_k} \right] + \sqrt{2} \, H_{\mathrm{D}6}^{-1/2} H_{\mathrm{NS}5} \, dz^2 + \frac{1}{\sqrt{2}} H_{\mathrm{D}6}^{1/2} H_{\mathrm{NS}5} \left(d\rho^2 + \rho^2 ds^2_{\tilde{S}^2}\right) \,, \\
F_{(2)} &= \frac{Q_{\mathrm{D}6}}{2} \, \text{vol}_{\tilde S^2} \,,  \qquad \qquad  e^{\Phi} = 2^{3/4} H_{\mathrm{D}6}^{-3/4} H_{\mathrm{NS}5}^{1/2} \,,\\
H_{(3)} &= -\partial_\rho H_{\mathrm{NS}5} \, \rho^2 \, dz \wedge \text{vol}_{\tilde S^2} + \frac12 \, H_{\mathrm{D}6} \, \partial_z H_{\mathrm{NS}5} \, \rho^2 \, d\rho \wedge \text{vol}_{\tilde S^2} \,, \\
 F_{(4)} &= 8 \, k\,Q_{\text{D}4} \, \text{vol}_{\text{AdS}_3}\wedge dz + 8 \, k \, Q_{\text{D}4} \, \text{vol}_{S^3/\mathbb{Z}_k} \wedge dz \,,\\
\end{split}
\end{equation}
with
\begin{equation}
\label{10d-motherbranesEOM_nh}
\nabla^2_{\mathbb{R}^3_\rho} H_{\mathrm{NS}5} + \frac12 H_{\mathrm{D}6} \,\partial_z^2 H_{\mathrm{NS}5} =0 \qquad  \text{and}  \qquad  H_{\mathrm{D}6} = \frac{Q_{\mathrm{D}6}}{\rho} \,.
\end{equation}
The parameters $Q_{\mathrm{D}4}$ and $Q_{\mathrm{D}6}$ represent, respectively, the D4-D2 defect charges and the D6 background charges. The latter arise from the KK'-monopoles in 11d (see Figure \ref{fig:TSTchain} or \cite{Faedo:2020nol} for a more detailed study).
This solution is related by T-duality along the Hopf-fibre direction of the $S^3/\mathbb{Z}_k$, that we have denoted by $\chi$, to the AdS$_3$ solution in Type IIB with $\xi=\pi/2$, obtained in section~\ref{AdS3pq5}. This is also the case for the underlying brane solutions before the near-horizon limit is taken. In turn, the Type IIB solution is related by S-duality to the Type IIB solution with $\xi=0$, obtained in section~\ref{D3NS5D5setup}. This relation is in agreement with the common M-theory origin of the two Type IIA  solutions T-dual to them. The relations among the different AdS$_3$ solutions in Type II and M-theory are depicted in Figure \ref{fig:TSTchain}. The underlying brane set-ups are depicted in Table \ref{Table:branesinmasslessIIA}.
\begin{table}[h!]
\renewcommand{\arraystretch}{1}
\scalebox{0.82}[0.8]{
\begin{tabular}{c||c c|c c c || c | c c c c}
 branes & $t$ & $x^1$ & $r$ & $\theta^{1}$ & $\theta^{2}$ & $z$ & $\rho$ & $\varphi^1$ & $\varphi^2$ & $\phi$ \\
\hline \hline
$\mrm{D}2$ & $\times$ & $\times$ & $-$ & $-$ & $-$ & $\times$ & $-$ & $-$ & $-$ & $-$ \\
$\mrm{D}6$ & $\times$ & $\times$ & $-$ & $-$ & $-$ & $\times$ & $\times$ & $\times$ & $\times$ & $\times$ \\
$\mrm{NS}5$ & $\times$ & $\times$ & $-$ & $-$ & $-$ & $-$ & $\times$ & $\times$ & $\times$ & $\times$ \\
$\mrm{D}4$ & $\times$ & $\times$ & $\times$ & $\times$ & $\times$ & $-$ & $-$ & $-$ & $-$ & $-$ \\
$\mrm{KK}'$ & $\times$ & $\times$ & $\times$  & $\times$ & $\times$ & $\times$ & $-$ & $-$ &$-$& $\text{ISO}$ \\
\end{tabular}
\quad
\begin{tabular}{c||c c|c c c c || c | c c c}
branes & $t$ & $x^1$ & $r$ & $\theta^{1}$ & $\theta^{2}$ & $\chi$ & $z$ & $\rho$ & $\varphi^1$ & $\varphi^2$ \\
\hline \hline
$\mrm{D}2$ & $\times$ & $\times$ & $-$ & $-$ & $-$ & $-$ &  $\times$ & $-$ & $-$ & $-$  \\
$\mrm{KK}$ & $\times$ & $\times$ & $-$ & $-$ & $-$ & $\mrm{ISO}$ & $\times$ & $\times$ & $\times$ & $\times$  \\
$\mrm{D}4$ & $\times$ & $\times$ &$-$ & $-$ & $-$ & $-$ & $-$ & $\times$ & $\times$ & $\times$  \\
$\mrm{NS}5$ & $\times$ & $\times$ & $\times$ & $\times$& $\times$  & $\times$ & $-$ & $-$ & $-$ & $-$  \\
$\mrm{D}6$ & $\times$ & $\times$ & $\times$ & $\times$ & $\times$ & $\times$ & $\times$ & $-$ & $-$ & $-$  \\
\end{tabular}
}
\caption{BPS/8 brane intersections underlying the AdS$_3\times S^2\times \tilde{S}^3/\mathbb{Z}_{k'}$ (left-side) and  AdS$_3\times S^3/\mathbb{Z}_k \times S^2$ (right-side) solutions to Type IIA of  \cite{Lozano:2019emq} and  \cite{Faedo:2020nol}. The intersections are T-duals to D3-NS5-D5 branes ending on NS5'-D5' background branes.} \label{Table:branesinmasslessIIA}
\end{table}
Note that the two aforementioned Type IIA string backgrounds define the simplest T-dual configurations of a plethora of interpolating $\mathrm{AdS}_3$ solutions in Type IIB at a generic $\xi$ angle, describing the compactifications of the original M-theory set-up on the $T^2$ parametrised by the Taub-NUT directions of the two 11d KK-monopoles. 


\section{Field theory interpretation}
\label{FT-interpretation}

In this section we construct quiver CFTs associated to the D3-brane boxes dual to our AdS$_3$ solutions. However, using that the $\xi=0$ solutions are T-dual to the AdS$_3\times S^2\times T^4/\mathbb{Z}_{k'}\times I_z$ solutions in \cite{Lozano:2019emq}, they should be dual to the same 2d CFTs, constructed in \cite{Lozano:2019jza,Lozano:2019zvg}, and we can avoid going through the detailed analysis of the field theory associated to brane boxes, which simplifies a lot our analysis. The associated quivers  consist on linear quivers along the $z$-direction that are repeated $k'$ times along the T-dual circle.  As recalled in the previous section, the solutions with $\xi=\pi/2$ are T-dual to the class of AdS$_3\times S^3/\mathbb{Z}_k\times S^2$ solutions to Type IIA constructed in \cite{Faedo:2020nol}. Therefore, our analysis in this section allows us to provide CFT duals to these solutions as well, by defining them globally in a specific manner that we discuss. As a non-trivial check of our proposal we show that the central charge matches the holographic result, connecting with our results in subsection \ref{holocc}.

\subsection{2d CFTs dual to the $\xi=0$ solutions}

\begin{figure}
\centering
\includegraphics[scale=0.6]{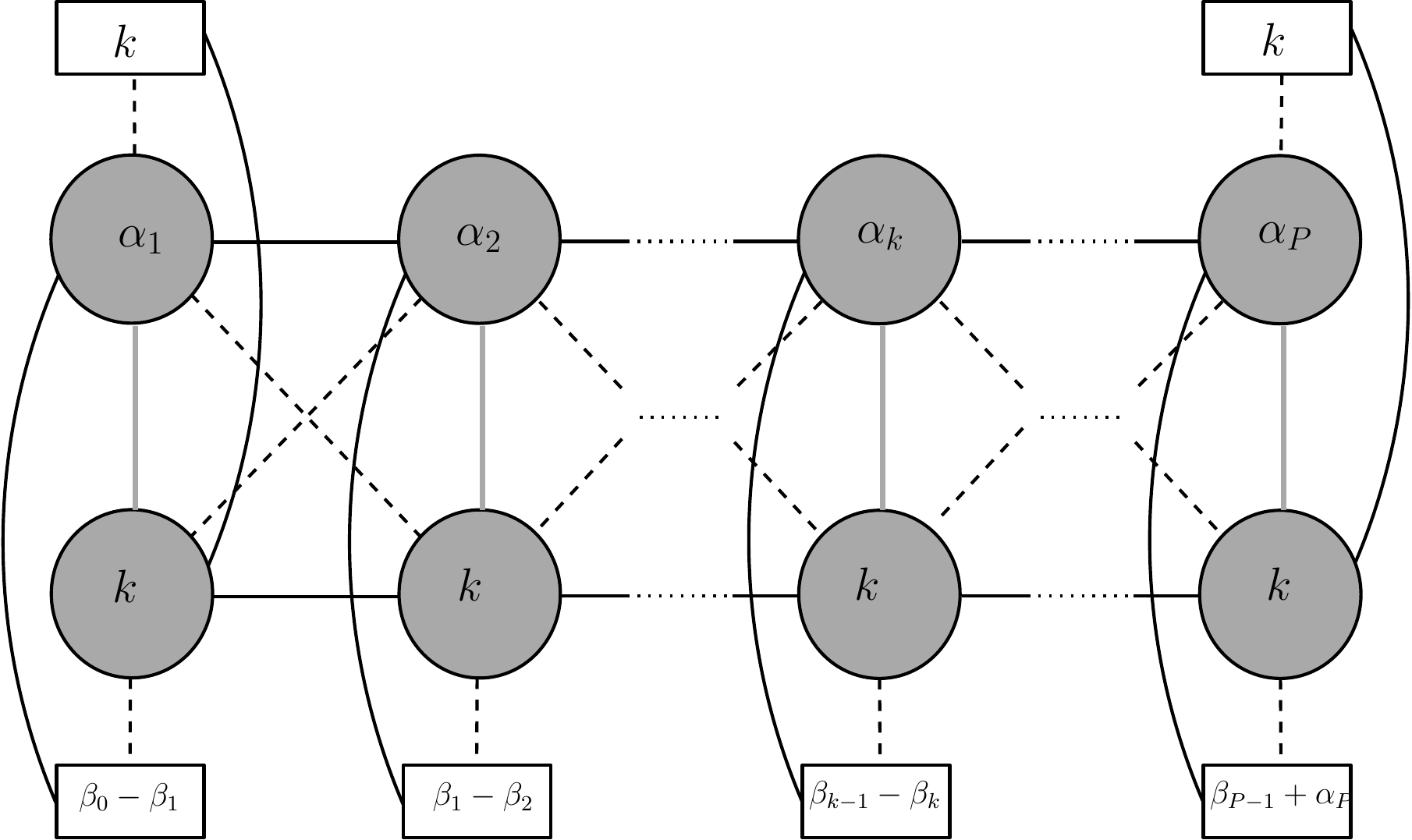}
\caption{Generic quiver field theory dual to the AdS$_3$ solutions in Type IIB with $\xi=0$.}
\label{quiver1}
\end{figure}

We start recalling the 2d CFTs dual to the solutions with $\xi=0$. Given that these solutions are T-dual to the AdS$_3\times S^2\times T^4/\mathbb{Z}_{k'}\times I_z$ solutions to Type IIA constructed in  \cite{Lozano:2019emq}, they are holographically dual to the same 2d CFTs. These CFTs were identified in \cite{Lozano:2019jza,Lozano:2019zvg} (see \cite{Lozano:2020bxo} for the restriction to the massless case)\footnote{Recall that our Type IIB configurations do not include D7-branes, so they are related to the solutions in  \cite{Lozano:2019emq} restricted to the massless case.}, for the case in which $H_{\mathrm{D}5'}$ is independent of the coordinates of the $T^4$. We have summarised the main properties of the construction in  \cite{Lozano:2019jza,Lozano:2019zvg,Lozano:2020bxo} in Appendix \ref{summary2dCFT}, to which the reader is referred for the details. Translated to Type IIB language, the quivers proposed therein contain gauge groups associated to D3 and D5 colour branes stretched between NS5-branes in the $z$ direction, as well as flavour groups associated to D5'-branes, orthogonal to them. In the $\phi$ direction, that parametrises the $S^1$, the D3-branes stretch between $k'$ NS5'-branes, repeating the configuration in each $z$ interval $k'$ times. The specific matter content guarantees gauge anomaly cancellation in 2d. We have depicted the IIB quivers for $k'=1$ in Figure \ref{quiver1}.
 They are in one to one correspondence with the quantised charges associated to the solutions, given by\footnote{
The $\mathrm{AdS}_3\times S^2 \times \tilde S^2 \times S^1$ Type IIB solutions given by \eqref{brane_metric_D3D5NS5D5NS5rotated_nh} depend on the parameters $Q_{\mathrm{D}3}, \, Q_{\mathrm{D}5}, \, Q_{\mathrm{NS}5'}$ and on the function $H_{\mathrm{D}5'}$. For the sake of clarity and to make manifest the links with the IIA solutions of Appendix~\ref{summary2dCFT}, we have fixed them as: $Q_{\mathrm{D}3}=1, \,Q_{\mathrm{D}5}=k, \,Q_{\mathrm{NS}5'}=k'$. Furthermore, we have fixed the constant $u$ of the IIA solution \eqref{profileh4sp2} as $u=Q_{\mathrm{D}5}$. In this way we get $H_{\mathrm{D}5'}=h_4$ under T-duality.}
 \begin{align}
N_{\mathrm{D}5}^{(j)} &= \frac{1}{(2\pi)^2}\int_{S^2\times S^1}{\hat F}_{(3)}= Q_{\mathrm{D}5}=k  \,, \label{D5charge}\\
N_{\mathrm{D}5'}^{(j)} &= \frac{1}{(2\pi)^2}\int_{I_\rho\times {\tilde S}^2}{\hat F}_{(3)}
= 2\pi\,Q_{\mathrm{NS}5'}^{-1} H_{\mathrm{D}5'}' = \beta_j\,, \label{D5'charge} \\
\begin{split}
N_{\mathrm{D}3}^{(j)} &= \frac{1}{(2\pi)^4}\int_{I_\rho\times {\tilde S}^2 \times S^2} {\hat F}_{(5)}
= Q_{\mathrm{NS}5'}^{-1}\biggl(H_{\mathrm{D}5'}-\Bigl(z-2\pi j\Bigr)H_{\mathrm{D}5'}'\biggr) = \alpha_j\,,\label{D3charge}
\end{split} \\
N_{\mathrm{NS}5}^{(j)} &= \frac{1}{(2\pi)^2}\int_{I_z\times S^2} H_{(3)} = 1 \label{NS5charge2}\, ,\\
N_{\mathrm{NS}5'}^{(j)} &= \frac{1}{(2\pi)^2}\int_{S^1\times {\tilde S}^2} H_{(3)} =  Q_{\mathrm{NS}5'}=k'  \label{NS5'charge}\, ,
\end{align}
 where the charges are computed in the $2\pi j\leq z \leq 2\pi (j+1)$ interval, as required by the enforcement of the condition $\frac{1}{4\pi^2}\oint_{S^2}B_{(2)}\in [0,1)$. Indeed, as discussed in Appendix \ref{summary2dCFT}, this implies that $B_{(2)}$ has to be defined as 
 \begin{equation}
 B_{(2)}=-\frac12  \Bigl(z-2\pi j\Bigr) \, \text{vol}_{S^2} + \frac12 Q_{\mathrm{NS}5'}\cos{\varphi_1} \, d\varphi_2 \wedge d\phi \, ,
\end{equation}
for $2\pi j\leq z \leq 2\pi (j+1)$. Moreover, in order to recover the quantised charges prior to the T-duality transformation, the integration in $\rho$ has to be regularised. This is best done in the formalism of Appendix~\ref{summary2dCFT}, where the $\rho$ coordinate comes from the rewriting of the $T^4$ as a fibration over a 3-sphere
\begin{equation} \label{rho-integration}
(2\pi)^4 = \mathrm{Vol}_{T^4} \equiv \mathrm{Vol}_{{\tilde S}^3} \int \rho^3 \, d \rho 
\end{equation}
where\footnote{To be coherent with the IIA notation of Appendix~\ref{summary2dCFT} we reparametrized the coordinate $\rho$ as $\rho \rightarrow Q_{\mathrm{NS}5'}^{1/2}\rho^{1/2}$.} $ds^2_{T^4}=d  \rho^2+ \rho^2 ds^2_{{\tilde S}^3}$.
 This condition defines the $\xi=0$ solutions globally in such a way that they are related by T-duality to the AdS$_3\times S^2\times T^4$ solutions constructed in \cite{Lozano:2019emq}.
 
The defining function $H_{\mathrm{D}5'}$ that gives rise to the final expressions in \eqref{D5charge}-\eqref{NS5'charge} is given by (see \eqref{profileh4sp2}),
\begin{equation} \label{profileh4sp}
Q_{\mathrm{NS}5'}^{-1} H_{\mathrm{D}5'}(z) =
\left\{
\begin{array}{ccl}
  \frac{\beta_0}{2\pi}z    && 0 \leq z \leq 2\pi\\[0.5em]
  \alpha_j+\frac{\beta_j}{2\pi}\bigl(z-2\pi j\bigr)    && 2\pi j \leq z \leq 2\pi (j+1) \,,  \quad  j=1,\dots,P-1 \\[0.5em]
  \alpha_P-\frac{\alpha_P}{2\pi}\bigl(z-2\pi P\bigr)    && 2\pi P \leq z \leq 2\pi (P+1) \,,
\end{array}
\right.
\end{equation}
where 
\begin{equation}
\alpha_j=\sum_{i=0}^{j-1}\beta_i\, .
\end{equation}
The choice of constants is imposed by continuity of the metric and dilaton. This choice implies discontinuities in the RR-sector, that are interpreted as generated by D5'-brane sources in the background. In order to describe well-defined, finite, 2d CFTs the $I_z$-direction needs to be globally defined. This is achieved requiring that $H_{\mathrm{D}5'}(z)$ vanishes at both ends of the $z$-interval, that are taken at $z=0,\,2\pi(P+1)$. This introduces extra D5'-branes in the configuration, as depicted in Figure \ref{quiver1}.
 
 \subsubsection{Central charge}
 
As summarised in Appendix \ref{summary2dCFT}, the central charge of 2d $(0,4)$ CFTs is computed from the anomaly of the U(1)$_R$ current, from where \cite{Putrov:2015jpa}
\begin{equation}
c_{CFT}=6\,(n_{hyp}-n_{vec})\, .
\end{equation}
This gives for the quiver CFTs depicted in Figure \ref{quiver1},
\begin{align}
n_{hyp} &= \sum_{j=1}^{P-1}\alpha_j\alpha_{j+1}+\sum_{j=1}^{P-1}\alpha_j(\beta_{j-1}-\beta_j)+\alpha_P(\beta_{P-1}+\alpha_P)+(P-1)k^2+k\sum_{j=1}^P\alpha_j +\alpha_P^2+2k^2=\nonumber\\
& =\sum_{j=1}^P\alpha_j^2+\sum_{j=1}^P\alpha_j\beta_{j-1}+(P+1)k^2+k\sum_{j=1}^P\alpha_j, \label{nhyp}\\
n_{vec} &= \sum_{j=1}^P \alpha_j^2 + P k^2.\label{nvec}
\end{align}
 The final result for the central charge is
\begin{equation} \label{cFT}
c_{CFT} = 6\, \Bigl(k\sum_{j=1}^P \alpha_j+ \sum_{j=1}^P\alpha_j\beta_{j-1}+k^2 \Bigr)=6\, \Bigl(Q_{\text{D}5}\sum_{j=1}^P \alpha_j+ \sum_{j=1}^P\alpha_j\beta_{j-1}+Q_{\mathrm{D}5}^2 \Bigr).
\end{equation}
We can compare now this expression with the value of the holographic central charge computed in subsection \ref{holocc}. Using the regularisation \eqref{rho-integration} for the integration in $\rho$ we find
\begin{equation}
c_{hol} = \frac{3}{\pi}\frac{1}{k'}\int dz\, h_8\, h_4 = \frac{3}{\pi}Q_{\mathrm{D}5} Q_{\mathrm{NS}5'}^{-1}\int dz \, H_{\mathrm{D}5'}.
\end{equation}
Substituting $H_{\mathrm{D}5'}$ as given by equation \eqref{profileh4sp}, we obtain
\begin{equation}
c_{hol}=6\, \Bigl(Q_{\mathrm{D}5}\sum_{j=1}^P \alpha_j\Bigr).
\end{equation}
In these calculations it is implicit that the $Q_{D5}$ and $Q_{NS5'}$ charges are smeared over intervals of length $2\pi$. However, in our case the D5-brane charge is smeared over a $z$ interval of length $2\pi P$. This introduces an extra factor of $P$ in both the field theory and holographic calculations{\footnote{In \cite{Lozano:2019zvg} it was shown that the $\text{CY}_2$ becomes of sub-stringy size and therefore the supergravity solution is no longer valid, unless the previous rescaling, $Q_{D5}\mapsto Q_{D5}P$, is performed.}. In turn, it is customary to scale the $\phi$ direction such that it runs between 0 and $2\pi k'$, such that one NS5' brane is created at each $\phi=2\pi$ position. In such case we must introduce an extra factor of $k'$ in both the field theory and holographic calculations. The final result of both calculations is then
\begin{equation} \label{ccfinal}
c_{CFT}=6\, Q_{\text{NS}5'}\Bigl(Q_{\mathrm{D}5}P\sum_{j=1}^P \alpha_j+ \sum_{j=1}^P\alpha_j\beta_{j-1}+Q_{\mathrm{D}5}^2P^2 \Bigr),
\end{equation}
and
\begin{equation}
c_{hol} = 6\,Q_{\mathrm{D}5} Q_{\mathrm{NS}5'} P\sum_{j=1}^P \alpha_j \,.
\end{equation}
In \cite{Lozano:2019jza,Lozano:2019zvg} (see also \cite{Lozano:2020txg}) a variety of examples of long linear quivers with sparse flavour groups and large ranks were considered in detail, with the conclusion that both expressions agree to leading order, when the background is a trustable description of the CFT. The reader is referred to these references for more details. Here we just present a simple example where this is shown explicitly. This is the quiver depicted in Figure \ref{example-quiver}.
\begin{figure}[h!]
\centering
\includegraphics[width=9cm]{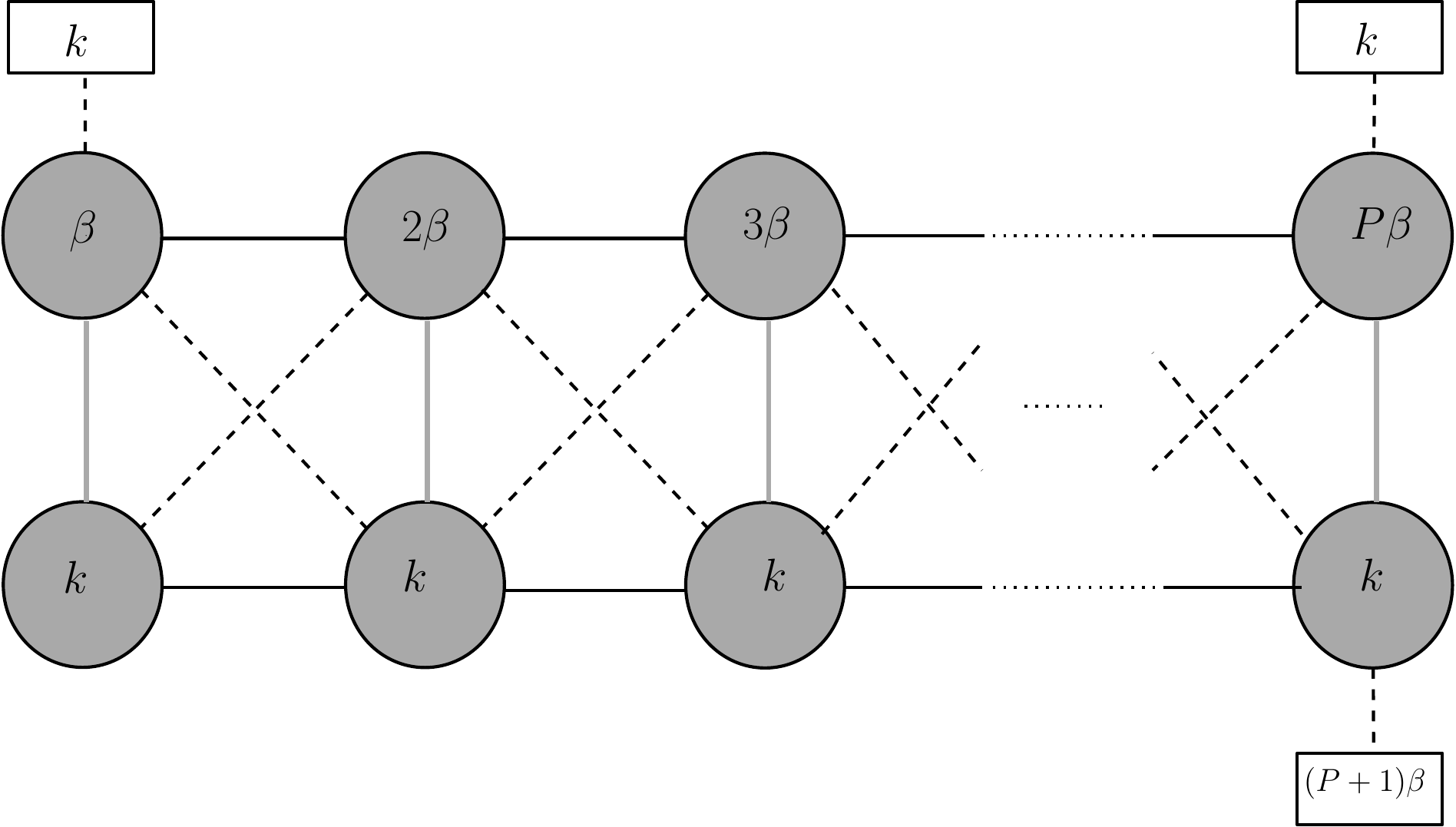}%
\caption{Example of 2d (0,4) quiver with gauge groups of linearly increasing ranks.}
\label{example-quiver}
\end{figure}
 In this case we have $\beta_j=\beta$ for $j=1,\dots,P-1$, and $\alpha_j=jP$ for $j=1,\dots,P$. The calculation of the field theory central charge gives
\begin{equation}
c_{CFT}=6\, Q_{\text{NS}5'}\Bigl(Q_{\text{D}5}P\, \frac{P(P+1)}{2}+ \beta\, \frac{P(P+1)}{2}+Q_{\text{D}5}^2P^2 \Bigr)\sim 3\, Q_{\text{NS}5'}Q_{\text{D}5}P^3,
\end{equation}
to leading order in $P$, in agreement with the holographic result
\begin{equation}
c_{hol}=6\, Q_{\text{NS}5'}Q_{\text{D}5}P\, \frac{P(P+1)}{2}\sim 3\, Q_{\text{NS}5'}Q_{\text{D}5}P^3\, .
\end{equation}

 \subsection{2d CFTs dual to the $\xi=\pi/2$ solutions}
 
Given the S-duality relation between the $\xi=0$ and the $\xi=\pi/2$ solutions, the quivers discussed in the previous section describe as well the 2d CFTs dual to the latter solutions. 
Indeed, as shown in \cite{Aharony:1997ju}, it is possible to read the CFT S-dual to a given CFT realised in a 5-brane web by  rotating the brane web by 90 degrees, with NS5 and D5 branes exchanged.  An example of this is depicted in Figure \ref{5dquivers}. 
In our case, given that our D3-brane box configuration is S-duality invariant, the same quivers described in the previous subsection are dual as well to the $\xi=\pi/2$ solutions. 
\begin{figure}
\centering
\includegraphics[scale=0.6]{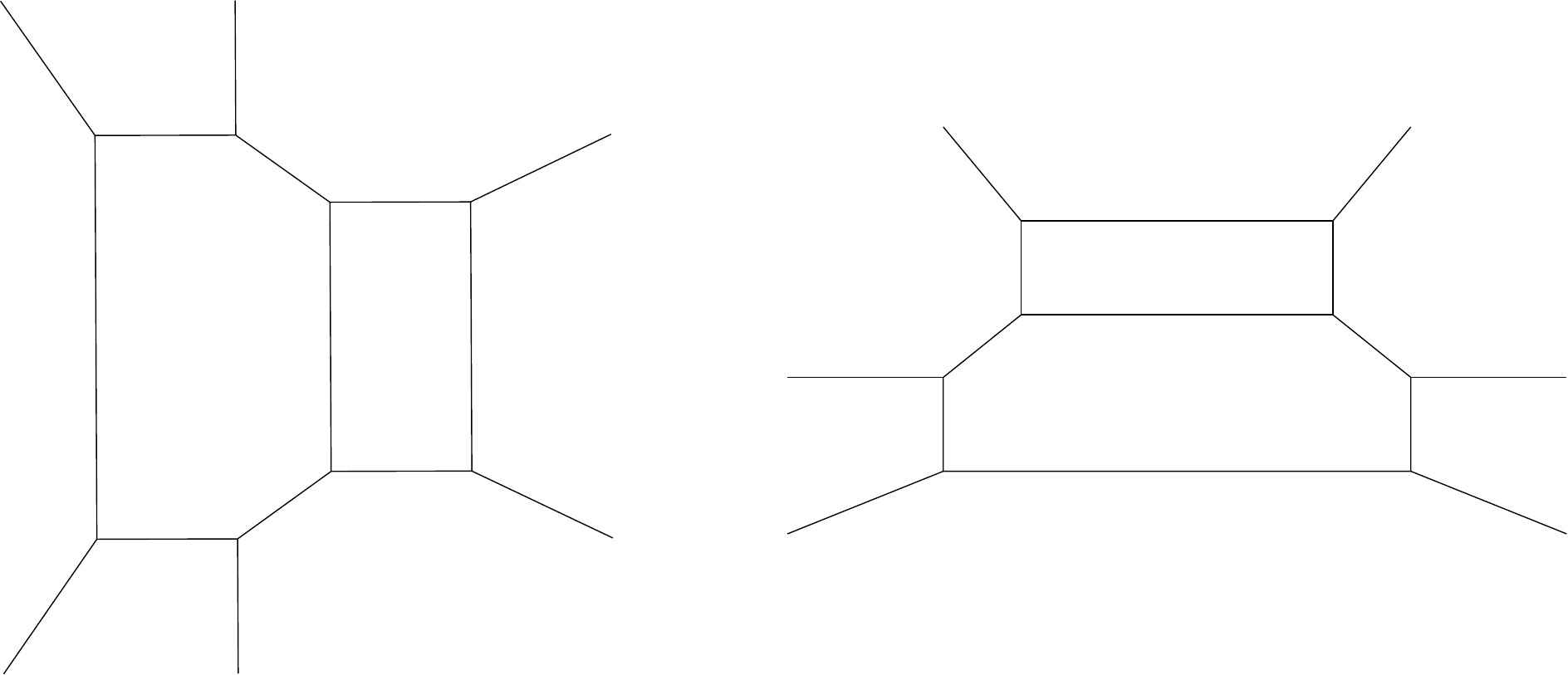}
\caption{Example of a 5d quiver CFT and its S-dual. In the left figure horizontal lines are associated to D5-branes and vertical lines to NS5-branes. It describes a 5d gauge theory with gauge group $\mathrm{SU}(2)\times \mathrm{SU}(2)$. In the right figure D5 and NS5 branes are exchanged, such that it describes a 5d gauge theory with gauge group $\mathrm{SU}(3)$ and two fundamental flavours, S-dual to the former.}
\label{5dquivers}
\end{figure} 
It is instructive however to 
show how this description arises in the strong coupling regime. 
In this regime the low-energy field content arises from open D1-strings living in D3 and NS5 colour branes stretched in $z$ between D5 branes, with extra NS5' flavour branes orthogonal to them. The quantisation of the open D1-strings gives rise in the strong coupling regime to the same field content as the quantisation of open F1-strings in the weak coupling regime. This is guaranteed by the S-duality invariance of Type IIB string theory. This can be realised explicitly through a careful computation of the quantised charges associated to the $\xi=\pi/2$ solution. This computation gives
\begin{align}
N_{\mathrm{NS}5}^{(j)} &= \frac{1}{(2\pi)^2}\int_{S^2\times S^1}H_{(3)} = Q_{\mathrm{D}5} \,, \label{NS5charge3}\\
N_{\mathrm{NS}5'}^{(j)} &= \frac{1}{(2\pi)^2}\int_{I_\rho\times {\tilde S}^2}H_{(3)} = 2\pi\,Q_{\mathrm{NS}5'}^{-1}\,H_{\mathrm{D}5'}' = \beta_j\,, \label{NS5'charge2} \\
N_{\mathrm{D}3}^{(j)} &= \frac{1}{(2\pi)^4}\int_{I_\rho\times {\tilde S}^2 \times S^2} {\hat F}_{(5)} = Q_{\mathrm{NS}5'}^{-1}\biggl(H_{\mathrm{D}5'}-\Bigl(z-2\pi j\Bigr) H_{\mathrm{D}5'}'\biggr) = \alpha_j\,,\label{D3charge2} \\
N_{\mathrm{D}5}^{(j)} &= \frac{1}{(2\pi)^2}\int_{I_z\times S^2} \hat{F}_{(3)} = 1 \label{D5charge2}\, ,\\
N_{\mathrm{D}5'}^{(j)} &= \frac{1}{(2\pi)^2}\int_{S^1\times {\tilde S}^2} \hat{F}_{(3)} = Q_{\mathrm{NS}5'} \label{D5'charge2}\, .
\end{align}
From these expressions, and using open D1-strings as the light degrees of freedom of our theory, we can easily reproduce the quivers depicted in Figure \ref{quiver1}, through an analysis identical to the one performed in the previous subsection. Let us now explain the origin of these quantised charges.

The expressions \eqref{NS5charge3}-\eqref{D5'charge2} are computed from the Page fluxes associated to the $\xi=\pi/2$ solution, given by (\ref{brane_metric_D3D5NS5D5NS5rotated_nh}). They are S-dual to the quantised charges of the $\xi=0$ solutions, given by equations (\ref{D5charge})-(\ref{NS5'charge}). In order to achieve this we have taken due care of the large gauge transformations of the $C_{(2)}$ RR-potential.
As we have discussed, in the $\xi=0$ solutions a large gauge transformation of gauge parameter $j$ needs to be performed for $2\pi j\leq z \leq 2\pi (j+1)$ in order to satisfy that $\frac{1}{4\pi^2}\oint_{S^2}B_{(2)}$ lies in the fundamental region $[0,1)$. This condition becomes $\frac{1}{4\pi^2}\oint_{S^2}C_{(2)}\in [0,1)$ after S-duality. Note that even if this condition cannot arise in perturbative string theory, it arises non-perturbatively when the fundamental degrees of freedom come from open D1-strings. This gives
\begin{equation}
C_{(2)} = \frac{1}{2} \Bigl(z-2\pi j\Bigr) \, \text{vol}_{S^2}-\frac{1}{2}Q_{\mathrm{NS}5'}\cos{\varphi_1} \, d\varphi_2 \wedge d\phi\, ,
\end{equation}
for $2\pi j\leq z \leq 2\pi (j+1)$. This must be taken into account in the calculation of the quantised charges, where $z$ must be integrated in the $2\pi j\leq z \leq 2\pi (j+1)$ interval. Moreover, in the absence of D7-branes, the RR 5-form Page flux transforms under S-duality as
\begin{equation}
{\hat F}_{(5)}=F_{(5)}-F_{(3)}\wedge B_{(2)} \longrightarrow {\hat F}_{(5)}=F_{(5)} + H_{(3)}\wedge C_{(2)}.
\end{equation}
This is the expression for the ${\hat F}_{(5)}$ Page flux used in equation \eqref{D3charge2} in order to reproduce the D3-brane quantised charge. Taking these features into account and the regularisation prescription \eqref{rho-integration} for the $\rho$-integrals, we finally reproduce the quantised charges given by (\ref{NS5charge3})-(\ref{D5'charge2}), realising explicitly the S-duality invariance of Type IIB string theory.

\subsection{2d CFTs dual to the AdS$_3\times S^3/\mathbb{Z}_{k}\times S^2\times I_\rho\times I_z$ solutions to Type~IIA}

As we discussed in the previous section, the $\xi=\pi/2$ solutions to Type IIB whose dual CFTs we have just identified are T-dual to the class of solutions to Type IIA supergravity found in \cite{Faedo:2020nol}. These solutions are described by AdS$_3\times S^3/\mathbb{Z}_{k}\times S^2$ geometries fibred over two intervals. 

One can easily check that upon T-duality along the $\phi$ direction, the $S^2$ and the $S^1$ of the solutions defined by (\ref{brane_metric_D3D5NS5D5NS5rotated_nh}) build up an $S^3$  modded out by $\mathbb{Z}_k$, and give rise to the AdS$_3\times S^3/\mathbb{Z}_{k}\times S^2\times I_\rho\times I_z$ solutions to Type IIA given by (\ref{brane_metric_D2D4KKNS5D6_nh}), found in  \cite{Faedo:2020nol}. This class  of solutions must thus be holographically dual to the same 2d CFTs dual to the $\xi=\pi/2$ solutions. These CFTs have been identified in the previous subsection for $H_{\mathrm{D}5'}$ independent of the $\rho$ direction and with the specific dependence on $z$ defined by equation (\ref{profileh4sp}). Therefore, these quivers are dual to the same subclass of AdS$_3\times S^3/\mathbb{Z}_{k}\times S^2\times I_\rho\times I_z$ solutions to Type IIA. In this case the nodes of the quivers are associated to D2 and D4 branes stretched between NS5-branes along the Hopf-fibre direction of the $S^3/\mathbb{Z}_{k}$, that we have denoted by $\chi$ in the previous section.

A clash seems to arise at this point between the previous CFT description and the 2d dual CFTs proposed in  \cite{Faedo:2020nol}, holographically dual as well to the AdS$_3\times S^3/\mathbb{Z}_{k}\times S^2\times I_\rho\times I_z$ solutions. The way out to this apparent puzzle is that the solutions studied in \cite{Faedo:2020nol} are defined by a function $H_{\mathrm{D}5'}$ (or $h_4$ in the notation of \cite{Faedo:2020nol})  that depends on both intervals $z$ and $\rho$ in a very specific manner, that allows to complete them globally by an AdS$_7$ solution in the UV. This solution is defined by a foliation of AdS$_7\times S^2$ over an interval\footnote{More specifically, it is the restriction to the massless case of the general class of AdS$_7$ solutions to massive Type IIA constructed in \cite{Apruzzi:2013yva}.}. The interval arises as a combination of the $z$ and $\rho$ coordinates of the AdS$_3$ solutions, while the $S^2$ is identified with the $S^2$ in the internal space of the AdS$_3$ geometries. In turn, the AdS$_3$ subspace combines with the $S^3/\mathbb{Z}_k$ and another combination of the $z$ and $\rho$ directions to build up an AdS$_7$ space asymptotically. Closing the interval present in the internal space of the AdS$_7$ solution with D6-branes, as done in \cite{Apruzzi:2013yva}, a globally completed solution can then be defined. In this global completion the 2d CFT is deconstructed into a 6d CFT in the UV, while it remains two-dimensional at lower energy scales, describing a surface defect CFT embedded within the 6d CFT. Explicit quiver CFTs realising this feature explicitly were constructed in~\cite{Faedo:2020nol}.

Instead, the 2d quiver CFTs that we have proposed in this section are dual to the solutions defined by $H_{\mathrm{D}5'}(z)$ given  by equation \eqref{profileh4sp}. Being $H_{\mathrm{D}5'}$ $\rho$-independent, the solutions cannot be completed globally with an AdS$_7$ space and one needs to resort to an alternative mechanism that renders a finite 2d dual CFT. We have used the T+S+T duality transformation that relates these solutions to the AdS$_3 \times S^2\times \text{CY}_2$ solutions constructed in \cite{Lozano:2019emq} to propose such mechanism. Our analysis shows that the general class of AdS$_3\times S^3/\mathbb{Z}_{k}\times S^2\times I_\rho\times I_z$ solutions to Type IIA found in  \cite{Faedo:2020nol} admits rather different CFT realisations depending on the explicit form of the $H_{\mathrm{D}5'}$ function that defines the solutions, together with the prescription taken to define them globally.

\section{Surface defects within AdS$_6$ in Type IIB}
\label{AdS6-defects}

In this section we extend the analysis of section \ref{AdS3solutions} by including D7-branes. We show that a subclass of these solutions asymptotes locally to the AdS$_6\times S^2\times \Sigma_2$ solution in the general class of \cite{DHoker:2016ujz,DHoker:2017mds,DHoker:2017zwj} that arises upon T-duality from the Brandhuber-Oz solution \cite{Brandhuber:1999np}. 

The T-dual of the Brandhuber-Oz solution was obtained in \cite{Cvetic:2000cj}, and further discussed in \cite{Lozano:2012au,Lozano:2018pcp}. The Type IIB interpretation is given in terms of a D5-NS5-D7 brane intersection that reproduces, in its near-horizon, an AdS$_6\times S^2\times \Sigma_2$ solution in the general classification in \cite{DHoker:2016ujz,DHoker:2017mds,DHoker:2017zwj}, with $\Sigma_2$ an annulus. This solution represents the only known explicit solution where $\Sigma_2$ is an annulus in this general classification.
 
In this section we put on top of the aforementioned intersection extra D3-NS5-D5 branes. This breaks the isometries of the 5d Minkowski spacetime in the D5-NS5-D7 brane intersection to $\mathrm{SO(2,1)}\times \mathrm{SO}(3)$. The first one is the group of isometries of the 2d Minkowski spacetime, while $\mathrm{SO}(3)$ is an R-symmetry group. The near-horizon regime associated to this bound state extends the class of  $\ma N=(0,4)$ $\mathrm{AdS}_3\times S^2 \times \tilde S^2 \times S^1\times I_z\times I_\rho$ vacua of Type IIB string theory obtained in section \ref{AdS3solutions} to include D7-branes. These vacua are T-dual to the $\mathrm{AdS}_3\times S^2 \times T^4/\mathbb{Z}_{k'} \times I$ solutions to massive IIA in \cite{Lozano:2019emq}, with the T-duality taking place along the Hopf-fibre direction of the 3-sphere contained in the $T^4/\mathbb{Z}_{k'}$. Within this class of solutions we identify the ones that asymptote locally to $\mrm{AdS}_6$, and thus allow for a defect interpretation within a 5d fixed point theory. For this purpose we relate the $\mathrm{AdS}_3$ backgrounds with a 6d domain wall reproducing asymptotically locally $\mathrm{AdS}_6$, found in \cite{Dibitetto:2018iar}. In doing this we make crucial use of the compactification of Type IIB around the $\mrm{AdS}_6\times S^2\times \Sigma_2$ vacua in \cite{DHoker:2016ujz,DHoker:2017mds,DHoker:2017zwj} constructed in \cite{Hong:2018amk}.

\subsection{The T-dual of Brandhuber-Oz from D5-NS5-D7 branes}

In this section we briefly explain the brane construction that underlies the $\mathrm{AdS}_6\times S^2 \times \Sigma_2$ vacuum T-dual to the Brandhuber-Oz solution. This will be useful when we include extra defect branes in the next subsection. 

It is well-known that in order to reproduce the $\mathrm{AdS}_6\times S^4$ vacuum of massive Type IIA \cite{Brandhuber:1999np} from a brane set-up, one has to include KK-monopoles, introducing an A-type singularity in the space transverse to the D4-D8 branes \cite{Cvetic:2000cj}. T-dualising to Type IIB along the Taub-NUT direction of the monopoles gives rise to the D5-NS5-D7 brane system depicted in Table \ref{Table:branesIIBAdS6}.
\begin{table}[http!]
\renewcommand{\arraystretch}{1}
\begin{center}
\scalebox{1}[1]{
\begin{tabular}{c||c cc c c || c | c c c | c}
 branes & $t$ & $x^1$ & $x^2$ & $x^3$ & $x^4$ & $z$ & $\rho$ & $\varphi^1$ & $\varphi^2$ & $\phi$ \\
\hline \hline
$\mrm{D}5$ & $\times$ & $\times$ & $\times$ & $\times$ & $\times$ & $-$ & $-$ & $-$ & $-$ & $\times$ \\
$\mrm{NS}5$ & $\times$ & $\times$ & $\times$  & $\times$ & $\times$ & $\times$ & $-$ & $-$ &$-$& $-$ \\
$\mrm{D}7$ & $\times$ & $\times$ & $\times$  & $\times$ & $\times$ & $-$ & $\times$ & $\times$ &$\times$& $-$ \\
\end{tabular}
}
\end{center}
\caption{Brane picture underlying the $\mathrm{AdS}_6\times S^2 \times \Sigma_2$ solution T-dual to Brandhuber-Oz. The system is $\mrm{BPS}/4$.} \label{Table:branesIIBAdS6}
\end{table}
Following the discussion in \cite{Cvetic:2000cj}, we now derive the $\mathrm{AdS}_6\times S^2 \times \Sigma_2$ geometry that arises at the near-horizon of this brane intersection, by providing the explicit parametrisation in which $\mathrm{AdS}_6$ is manifest. The background that arises is the T-dual of the Brandhuber-Oz solution, with the T-duality taking place along the Hopf-fibre direction of the half 3-sphere.

We start considering the following metric
\begin{equation}
\label{brane_metric_D5NS5D7}
\begin{split}
d s_{10}^2&=H_{\mathrm{D}7}^{-1/2}H_{\mathrm{D}5}^{-1/2}ds^2_{\mathbb{R}^{1,4}} +H_{\mathrm{D}7}^{1/2}H_{\mathrm{D}5}^{1/2}dz^2+H_{\mathrm{D}7}^{1/2}H_{\mathrm{D}5}^{-1/2}H_{\mathrm{NS}5}d\phi^2+H_{\mathrm{D}7}^{-1/2}H_{\mathrm{D}5}^{1/2}H_{\mathrm{NS}5}\left(d\rho^2+\rho^2ds^2_{S^2}\right),
\end{split}
\end{equation}
with gauge potentials and dilaton given by
\begin{equation}
\begin{split}\label{brane_potentials_D5NS5D7}
&B_{(6)}=H_{\mathrm{D}7} H_{\mathrm{NS}5}^{-1}\,\text{vol}_{\mathbb{R}^{1,4}}\wedge dz\,,\\ \vspace{0.4cm}
&C_{(6)}=H_{\mathrm{D}5}^{-1}\,\text{vol}_{\mathbb{R}^{1,4}}\wedge d\phi\,,\\
&C_{(8)}=H_{\mathrm{NS}5} H_{\mathrm{D}7}^{-1}\rho^2\,\text{vol}_{\mathbb{R}^{1,4}}\wedge d\rho \wedge \text{vol}_{S^2}\,,\\
&e^{\Phi}=H_{\mathrm{D}7}^{-1} H_{\mathrm{NS}5}^{1/2} H_{\mathrm{D}5}^{-1/2}\,.
\end{split}
\end{equation}
A simple solution can be obtained by assuming the smearing of the NS5 and D7 branes along the $\phi$ direction, i.e.\ $H_{\mathrm{NS}5}=H_{\mathrm{NS}5}(\rho)$ and $H_{\mathrm{D}7}=H_{\mathrm{D}7}(z)$. With this prescription the fluxes take the form 
\begin{equation}
\begin{split}\label{fluxes_D5NS5D7}
  &F_{(1)}=\partial_z H_{\mathrm{D}7}d\phi\\
&H_{(3)}=\partial_\rho H_{\mathrm{NS}5}\,\rho^2\text{vol}_{S^2} \wedge d\phi\,,\\
  &F_{(3)}=-H_{\mathrm{D}7}\,\partial_\rho H_{\mathrm{D}5}\,\rho^2dz\wedge\text{vol}_{S^2} +H_{\mathrm{NS}5}\,\partial_zH_{\mathrm{D}5}\,\rho^2 d\rho\wedge\text{vol}_{S^2}\,.\\
 \end{split}
\end{equation}
The equations of motion and Bianchi identities collapse to the following equations
\begin{equation}\label{NS5D5D7EOMS}
\nabla^2_{\mathbb{R}^3_\rho} H_{\mathrm{D}5} +H_{\mathrm{D}7}^{-1} H_{\mathrm{NS}5} \, \partial_z^2 H_{\mathrm{D}5}=0  \qquad  \text{and}  \qquad  \nabla^2_{\mathbb{R}^3_\rho} H_{\mathrm{NS}5} = 0  \,,
\end{equation}
with $H_{\mathrm{D}7}$ such that $\partial_z^2 H_{\mathrm{D}7}=0$.
A particular solution for the equations \eqref{NS5D5D7EOMS} is given by \cite{Cvetic:2000cj},
\begin{equation}\label{HfuncAdS6}
 H_{\mathrm{D}5}=1+\frac{Q_{\mathrm{D}5}}{\left(4Q_{\mathrm{NS}5}\rho+\frac49 Q_{\mathrm{D}7}z^3 \right)^{5/3}}\,,\qquad H_{\mathrm{NS}5}=\frac{Q_{\mathrm{NS}5}}{\rho}\,, \qquad H_{\mathrm{D}7}=Q_{\mathrm{D}7}z\,,
\end{equation}
where $Q_{\mathrm{D}5}$, $Q_{\mathrm{D}7}$ and $Q_{\mathrm{NS}5}$ are related to the quantised charges of the D5, D7 and NS5 branes.

The best system of coordinates to study the near-horizon limit of this background is obtained by combining the coordinates $(z, \rho)$ as 
\begin{equation}\label{coordAdS6}
 z=Q_{\mathrm{NS}5}^{1/3}\,\zeta\,s^{2/3}\,,\qquad \rho= \frac{1}{9}Q_{\mathrm{D}7}\,\zeta^{3}\,c^{2}\,,
\end{equation}
with $s=\sin\alpha$ and $c=\cos\alpha$. Expressing solution \eqref{HfuncAdS6} in terms of these new coordinates, the near-horizon regime is obtained in the limit $\zeta\to0$, when one can neglect the 1 factor in $H_{\mathrm{D}5}$. In this limit the 10d background takes the form\footnote{We rescaled the 5d worldvolume coordinates as $(t,x^i)\mapsto\ell^{-2}\,Q_{\mathrm{NS}5}^{-1/3}(t,x^i)$.}
\begin{equation}
 \begin{split}\label{brane_metric_D5NS5D7_nh}
  &\ell^2\,ds^2_{10}=s^{-1/3}\left[ds^2_{\text{AdS}_6}+9^{-1}c^2ds^2_{S^2}+\frac49 d\alpha^2+9\ell^4 Q_{{\scriptsize \mrm{NS}5}}^2\,c^{-2}s^{2/3}d\phi^2   \right]\,,\\
  &H_{(3)}=-Q_{\mathrm{NS}5}\text{vol}_{S^2}\wedge d\phi \,, \qquad e^{\Phi}=\frac{2^{5/3}}{3^{2/3}}\,\frac{Q_{\mathrm{NS}5}}{Q_{\mathrm{D}7}^{2/3}Q_{\mathrm{D}5}^{1/2}}\,c^{-1}s^{-2/3}\,,\\
  & F_{(1)}=Q_{\mathrm{D}7}d\phi\,,\\
  &F_{(3)}=\frac{5}{2^{7/3}3^{2/3}}\,\frac{Q_{\mathrm{D}7}^{1/3}Q_{\mathrm{D}5}}{Q_{\mathrm{NS}5}}\,c^3s^{1/3} \text{vol}_{S^2}\wedge d\alpha\,,
 \end{split}
\end{equation}
where $\ell^2=\left(\frac49\right)^{5/6}\,Q_{\mathrm{D}7}^{1/3}Q_{\mathrm{D}5}^{-1/2}$.
This background preserves 16 supercharges and, as shown in \cite{Lozano:2018pcp}, belongs to the general classification in \cite{DHoker:2016ujz,DHoker:2017mds,DHoker:2017zwj} for a 
Riemann surface $\Sigma_2$ with the topology of an annulus, parametrised by the coordinates $(\alpha,\phi)$. We would like to stress that the change of coordinates given by equation \eqref{coordAdS6} plays a crucial role in the derivation of the $\mathrm{AdS}_6$ background \eqref{brane_metric_D5NS5D7_nh}  from the brane solution.

\subsection{The brane set-up}

In the previous section we showed how to reproduce the T-dual of the Brandhuber-Oz solution as the near-horizon limit of the D5-NS5-D7 brane set-up depicted in Table \ref{Table:branesIIBAdS6}. In this subsection we intersect the aforementioned bound state with extra D3-NS5-D5 branes. The resulting brane set-up is depicted in Table \ref{Table:branesIIBD7}. The intersection with these additional branes breaks the isometries of $\mathrm{AdS}_6$ to those of $\mathrm{AdS}_3$ and reduces the number of supercharges to 8, organised in $\ma N=(0,4)$ Killing spinors with SU(2) R-symmetry.
\begin{table}[http!]
\renewcommand{\arraystretch}{1}
\begin{center}
\scalebox{1}[1]{
\begin{tabular}{c||c c|c c c || c | c c c c}
 branes & $t$ & $x^1$ & $r$ & $\theta^{1}$ & $\theta^{2}$ & $z$ & $\rho$ & $\varphi^1$ & $\varphi^2$ & $\phi$ \\
\hline \hline
$\mrm{D}5'$ & $\times$ & $\times$ & $\times$ & $\times$ & $\times$ & $-$ & $-$ & $-$ & $-$ & $\times$ \\
$\mrm{NS}5'$ & $\times$ & $\times$ & $\times$  & $\times$ & $\times$ & $\times$ & $-$ & $-$ &$-$& $-$ \\
$\mrm{D}7$ & $\times$ & $\times$ & $\times$  & $\times$ & $\times$ & $-$ & $\times$ & $\times$ &$\times$& $-$ \\
$\mrm{D}3$ & $\times$ & $\times$ & $-$ & $-$ & $-$ & $\times$ & $-$ & $-$ & $-$ & $\times$ \\
$\mrm{NS}5$ & $\times$ & $\times$ & $-$ & $-$ & $-$ & $-$ & $\times$ & $\times$ & $\times$ & $\times$ \\
$\mrm{D}5$ & $\times$ & $\times$ & $-$ & $-$ & $-$ & $\times$ & $\times$ & $\times$ & $\times$ & $-$ \\
\end{tabular}
}
\end{center}
\caption{Brane picture underlying the intersection of the D5'-NS5'-D7 branes depicted in Table \ref{Table:branesIIBAdS6} with extra D3-NS5-D5 branes. The system is $\mrm{BPS}/8$.} \label{Table:branesIIBD7}
\end{table}

The string background we are interested in can be obtained by plugging the brane solution \eqref{brane_metric_D3D5NS5D5NS5} on a D7 background.
Therefore, we consider the metric
\begin{equation}
\label{brane_metric_D3D5NS5D5NS5D7}
\begin{split}
d s_{10}^2 &= H_{\mathrm{D}7}^{-1/2} H_{\mathrm{D}5'}^{-1/2} \left[H_{\mathrm{D}5}^{-1/2} H_{\mathrm{D}3}^{-1/2}\,ds^2_{\mathbb{R}^{1,1}}+H_{\mathrm{D}5}^{1/2} H_{\mathrm{D}3}^{1/2}  H_{\mathrm{NS}5} \bigl(dr^2+r^2ds^2_{S^2}\bigr) \right] \\
&+H_{\mathrm{D}7}^{1/2} H_{\mathrm{D}5'}^{1/2} H_{\mathrm{NS}5} H_{\mathrm{D}5}^{-1/2} H_{\mathrm{D}3}^{-1/2}dz^2+H_{\mathrm{D}7}^{1/2} H_{\mathrm{D}5'}^{-1/2} H_{\mathrm{NS}5'} H_{\mathrm{D}5}^{1/2} H_{\mathrm{D}3}^{-1/2}d\phi^2\\
&+H_{\mathrm{D}7}^{-1/2} H_{\mathrm{D}5'}^{1/2} H_{\mathrm{NS}5'} H_{\mathrm{D}5}^{-1/2} H_{\mathrm{D}3}^{1/2} \bigl(d\rho^2+\rho^2ds^2_{\tilde S^2}\bigr)\,.
\end{split}
\end{equation}
We require that the D3-NS5-D5 branes are fully localised within the worldvolume of the NS5'-D5'-D7 branes. This assumption implies the smearing of the corresponding warp factors, i.e.\ 
$H_{\mathrm{D}5}=H_{\mathrm{D}5}(r)$, $H_{\mathrm{NS}5}=H_{\mathrm{NS}5}(r)$ and $H_{\mathrm{D}3}=H_{\mathrm{D}3}(r)$. Concerning the charge distributions within the NS5'-D5'-D7 brane intersection, we consider the ones that give rise to the $\mrm{AdS}_6$ background of the previous subsection, namely $H_{\mathrm{NS}5'}=H_{\mathrm{NS}5'}(\rho)$, $H_{\mathrm{D}7}=H_{\mathrm{D}7}(z)$ and the D5' branes fully localised in their transverse space.
We introduce the following gauge potentials and dilaton associated to this particular charge distribution,
\begin{equation}
\begin{split}\label{brane_potentials_D3NS5D5NS5D5D7}
B_{(6)} &= H_{\mathrm{D}7}H_{\mathrm{NS}5}H_{\mathrm{D}5}H_{\mathrm{NS}5'}^{-1}r^2\,\text{vol}_{\mathbb{R}^{1,1}}\wedge dr \wedge \text{vol}_{S^2}\wedge dz \\
& +H_{\mathrm{D}7}H_{\mathrm{NS}5'}H_{\mathrm{D}5'}H_{\mathrm{NS}5}^{-1}\rho^2\text{vol}_{\mathbb{R}^{1,1}} \wedge d\rho \wedge \text{vol}_{\tilde S^2}\wedge d\phi,\\ \vspace{0.4cm}
C_{(8)} &= H_{\mathrm{D}3}H_{\mathrm{NS}5}H_{\mathrm{NS}5'}H_{\mathrm{D}7}^{-1}\,r^2\rho^2\,\text{vol}_{\mathbb{R}^{1,1}}\wedge dr \wedge \text{vol}_{S^2}\wedge d\rho \wedge \text{vol}_{\tilde S^2}\,,\\
C_{(6)} &= H_{\mathrm{NS}5} H_{\mathrm{D}5} H_{\mathrm{D}5'}^{-1}\,r^2\,\text{vol}_{\mathbb{R}^{1,1}}\wedge dr \wedge \text{vol}_{ S^2}\wedge d\phi\\
&+H_{\mathrm{NS}5'} H_{\mathrm{D}5'} H_{\mathrm{D}5}^{-1}\,\rho^2\,\text{vol}_{\mathbb{R}^{1,1}}\wedge dz \wedge d\rho \wedge \text{vol}_{\tilde S^2}\,,\\
C_{(4)} &= H_{\mathrm{D}7} H_{\mathrm{D}3}^{-1}\,\text{vol}_{\mathbb{R}^{1,1}}\wedge dz\wedge d\phi\,,\\
e^{\Phi}&= H_{\mathrm{D}7}^{-1} H_{\mathrm{D}5'}^{-1/2} H_{\mathrm{NS}5'}^{1/2} H_{\mathrm{D}5}^{-1/2} H_{\mathrm{NS}5}^{1/2}\,.
\end{split}
\end{equation}
The fluxes can be worked out from the above gauge potentials,
\begin{equation}
\begin{split}\label{fluxes_D3NS5D5NS5D5D7}
H_{(3)} &= \partial_r H_{\mathrm{NS}5}\,r^2\text{vol}_{S^2} \wedge dz+\partial_\rho H_{\mathrm{NS}5'}\,\rho^2\text{vol}_{\tilde S^2} \wedge d\phi\,,\\
F_{(1)} &= H_{\mathrm{NS}5}^{-1} H_{\mathrm{D}5}\,\partial_z H_{\mathrm{D}7} \,d\phi\,,\\
F_{(3)} &= H_{\mathrm{D}7}\,\partial_r H_{\mathrm{NS}5}\,r^2\text{vol}_{S^2} \wedge d\phi-H_{\mathrm{D}7}\,\partial_\rho H_{\mathrm{D}5'}\,\rho^2dz\wedge\text{vol}_{\tilde S^2} \\
& +H_{\mathrm{D}3} H_{\mathrm{NS}5}^{-1} H_{\mathrm{NS}5'}\,\rho^2\,\partial_zH_{\mathrm{D}5'}\, d\rho\wedge\text{vol}_{\tilde S^2}\,,\\
F_{(5)} &= H_{\mathrm{D}7}\,\partial_r H_{\mathrm{D}3}^{-1}\,\text{vol}_{\mathbb{R}^{1,1}}\wedge dr\wedge dz \wedge d\phi+H_{\mathrm{D}5'} H_{\mathrm{NS}5'}\,r^2\rho^2\,\partial_rH_{\mathrm{D}3} \text{vol}_{S^2}\wedge \text{vol}_{\tilde S^2}\wedge d\rho\,.
\end{split}
\end{equation}
As for the intersection considered in section \ref{D3NS5D5setup}, the equations of motion and Bianchi identities of Type IIB supergravity decouple into two independent systems, describing, respectively, the D3-NS5-D5 and D5'-NS5'-D7 intersections. For the first intersection we get
\begin{equation} \label{D3D5NS5EOMS2}
\nabla^2_{\mathbb{R}^3_r} H_{\mathrm{D}3}=0 \qquad \text{with}\qquad  H_{\mathrm{D}5}=H_{\mathrm{NS}5}=H_{\mathrm{D}3}\,.
\end{equation}
This equation can be easily solved by means of harmonic functions,
\begin{equation}
H_{\mathrm{D}3}(r)=H_{\mathrm{NS}5}(r)=H_{\mathrm{D}5}(r)=1+\frac{Q_{\mathrm{D}3}}{r}\,.
\end{equation}
We point out that the number of independent parameters remains the same as in the background without D7 branes. In fact, the Bianchi identities for $F_{(3)}$ now relate the 5-brane charges $Q_{\mathrm{D}5}$, $Q_{\mathrm{NS}5}$ via the D7-branes flux $F_{(1)}$. In turn, the dynamics of the 
D5'-NS5'-D7 branes is completely fixed by the equations given by \eqref{NS5D5D7EOMS}, with $\partial_z^2 H_{\mathrm{D}7}=0$.

We now study this  10d geometry in the limit $r\to0$. This limit defines a class of solutions of Type IIB supergravity of the form\footnote{We rescaled the Minkowski coordinates as $(t,x^1)\mapsto2\,Q_{\mathrm{D}3}^{3/2}\,(t,x^1)$.}
\begin{equation}
\label{brane_metric_D3NS5D5NS5D5D7_nh}
\begin{split}
ds_{10}^2 &=  Q_{\text{D}3}^{2}H_{\mathrm{D}7}^{-1/2}H_{\mathrm{D}5'}^{-1/2}\left[4ds^2_{\text{AdS}_3} + ds^2_{S^2} \right] + H_{\mathrm{D}7}^{1/2}H_{\mathrm{D}5'}^{1/2}  dz^2+H_{\mathrm{D}7}^{1/2}H_{\mathrm{D}5'}^{-1/2}H_{\mathrm{NS}5'}  d\phi^2\\
&+H_{\mathrm{D}7}^{-1/2} H_{\mathrm{D}5'}^{1/2}H_{\mathrm{NS}5'} \left(d\rho^2 + \rho^2 ds^2_{\tilde{S}^2}\right) \,, \\
H_{(3)}&=-Q_{\text{D}3}\text{vol}_{S^2} \wedge dz+\partial_\rho H_{\mathrm{NS}5'}\,\rho^2\text{vol}_{\tilde S^2} \wedge d\phi\,,\qquad e^{\Phi}=H_{\mathrm{D}7}^{-1}H_{\mathrm{D}5'}^{-1/2}H_{\mathrm{NS}5'}^{1/2}\,,\\
F_{(1)}&=\partial_zH_{\mathrm{D}7} \,d\phi\,,\\
  F_{(3)}&=-Q_{\text{D}3}H_{\mathrm{D}7} \text{vol}_{S^2} \wedge d\phi-H_{\mathrm{D}7} \partial_\rho H_{\mathrm{D}5'}\,\rho^2dz\wedge\text{vol}_{\tilde S^2} + H_{\mathrm{NS}5'}\,\rho^2\,\partial_zH_{\mathrm{D}5'}\, d\rho\wedge\text{vol}_{\tilde S^2}\,,\\
 F_{(5)}&=8Q_{\text{D}3}^2H_{\mathrm{D}7} \,\text{vol}_{\text{AdS}_3}\wedge dz \wedge d\phi-Q_{\text{D}3}H_{\mathrm{D}5'}H_{\mathrm{NS}5'}\rho^2\,\text{vol}_{S^2}\wedge \text{vol}_{\tilde S^2}\wedge d\rho\,.
\end{split}
\end{equation}
The backgrounds \eqref{brane_metric_D3NS5D5NS5D5D7_nh} constitute a new class of $\ma N=(0,4)$ $\mrm{AdS}_3$ solutions to Type IIB string theory, for $H_{\mathrm{D}5'}$, $H_{\mathrm{NS}5'}$ and $H_{\mathrm{D}7}$ solutions of the equations
\begin{equation}\label{NS5D5D7EOMS2}
\nabla^2_{\mathbb{R}^3_\rho} H_{\mathrm{D}5} +H_{\mathrm{D}7}^{-1} H_{\mathrm{NS}5} \, \partial_z^2 H_{\mathrm{D}5}=0\,,\qquad  \nabla^2_{\mathbb{R}^3_\rho} H_{\mathrm{NS}5} = 0 \qquad  \text{and}  \qquad \partial_z^2 H_{\mathrm{D}7}=0 \,.
\end{equation}

In the next subsection we consider the particular solution defined by the equations in \eqref{HfuncAdS6}, giving rise in the UV to the $\mrm{AdS}_6$ vacuum T-dual to the Brandhuber-Oz solution.  We study the particular way in which the D3-NS5-D5 brane intersection breaks the isometries of the $\mrm{AdS}_6$ solution and gives rise to a surface defect at the locus of the intersection.

\subsection{Surface defects within the T-dual of Brandhuber-Oz}\label{defects}

In this section we show that the solutions in \eqref{brane_metric_D3NS5D5NS5D5D7_nh} can be given an interpretation as surface defects within the AdS$_6\times S^2 \times \Sigma_2$ solution T-dual to Brandhuber-Oz, given by \eqref{brane_metric_D5NS5D7_nh}. We reproduce this geometry from the uplift of a 6d domain wall characterised by an AdS$_3$ slicing and an asymptotic behaviour that reproduces locally the AdS$_6$ vacuum. This solution was found in~\cite{Dibitetto:2018iar} in the context of 6d $\mathcal{N}=(1,1)$ minimal gauged supergravity. See Appendix~\ref{6dsugra} for more details on this theory and its embedding in Type IIB. 

We consider the following 6d background\cite{Dibitetto:2018iar}
\begin{equation}
\begin{split}
\label{6dAdS3}
& ds^2_6 = e^{2U(\mu)} \left( 4\,ds^2_{\text{AdS}_3} + ds^2_{S^2} \right) + e^{2V(\mu)} d\mu^2 \,, \\
&\mathcal{B}_{(2)} = b(\mu) \, \text{vol}_{S^2} \,, \\
& X = X(\mu) \,.
\end{split}
\end{equation}
This background is described by the following set of BPS equations
\begin{equation}
\label{chargedDW6d}
U' = -2\,e^V f \,,  \qquad  X' = 2\,e^V X^2 \, D_X f \,,  \qquad  b' = \frac{e^{U+V}}{X^2} \,,
\end{equation}
together with the duality constraint\footnote{We take the 6d mass parameter $m=1$ and $g=\frac{3 }{\sqrt{2}}$ (see \cite{Dibitetto:2018iar}) to agree with the notation of the truncation formulas in \cite{Hong:2018amk}.}
\begin{equation}
\label{chargedDW6d1}
b = -e^{U} X \,,
\end{equation}
and the superpotential $f$ written in~\eqref{6dsuperpotential}. This flow preserves 8 real supercharges (BPS/2 in 6d). In order to obtain an explicit solution of \eqref{chargedDW6d}, a parametrisation of the 6d geometry needs to be chosen. One then obtains \cite{Dibitetto:2018iar},
\begin{equation} \label{chargedDWsol}
\begin{aligned}
e^{2U} &= 3^{-2/3}\,\left(\frac{\mu}{\mu^4-1}\right)^{2/3} \,,  &\qquad  e^{2V} &= \frac{16}{9}\,\frac{\mu^4}{(\mu^4-1)^2} \, \\
b &= -\frac{4}{3^{1/3}}\,\frac{\mu^{4/3}}{(\mu^4-1)^{1/3}} \,,  &\qquad  X &= \mu \,,
\end{aligned}
\end{equation}
with $\mu$ running between 0 and 1. 
In the regime $\mu \sim 1$ the 6d background is such that
\begin{equation}
R_{\,6} = -\frac{20}{3}\,g^2 + \mathcal{O}(1-\mu)^{2/3} \,,  \qquad  X = 1 + \mathcal{O}(1-\mu) \,,
\end{equation}
where $R_{\,6}$ is the 6d scalar curvature. The 2-form gauge potential gives sub-leading contributions in this limit. This implies that the asymptotic geometry for $\mu\to1$ is locally AdS$_6$. In the opposite limit, $\mu\to0$, the 6d background is manifestly singular. From a 10d point of view, this is due to the presence of the D3-D5-NS5 brane sources.

Let us consider now the uplift to Type IIB supergravity of the aforementioned 6d domain wall. Using the truncation formulas\footnote{Here the metric is given in the Einstein frame.} of \cite{Hong:2018amk}, summarised in \eqref{truncationansatz6d} and \eqref{10dfluxesto6d}, one obtains the following Type IIB $\mrm{AdS}_3$ background,
\begin{equation}
\begin{split}
\label{uplift6dDW}
ds_{10}^2 &= f_6^2 \, e^{2U} \left(4\,ds^2_{\text{AdS}_3} + ds^2_{S^2} \right) + f_6^2 \, e^{2V} d\mu^2 + f_2^2 \, ds_{\tilde{S}^2}^2 + 4N^2 \, dw d\bar{w} \,, \\
e^{-\Phi} &= \frac{2 (\alpha\bar\beta + \bar\alpha\beta)}{(\alpha+\bar\alpha)^2 - (\beta-\bar\beta)^2} \,,  \qquad  C_{(0)} = -i\,\frac{(\alpha^2 - \bar\alpha^2) - (\beta^2 - \bar\beta^2)}{(\alpha+\bar\alpha)^2 - (\beta-\bar\beta)^2} \,, \\
B_{(2)} &= i\,\frac{c_6}{9} \biggl[ \frac{\mathcal{C}-\bar{\mathcal{C}}}{\mathcal{D}} - 3(\mathcal{A}_+ + \bar{\mathcal{A}}_- - \bar{\mathcal{A}}_+ - \mathcal{A}_-) \biggr] \, \text{vol}_{\tilde{S}^2} + c_6 (\alpha + \bar\alpha) \, b \, \text{vol}_{S^2} \,, \\
C_{(2)} &= \frac{c_6}{9} \biggl[ \frac{\mathcal{C}+\bar{\mathcal{C}}}{\mathcal{D}} - 3(\mathcal{A}_+ + \bar{\mathcal{A}}_- + \bar{\mathcal{A}}_+ + \mathcal{A}_-) \biggr] \, \text{vol}_{\tilde{S}^2} - i\,c_6 (\alpha - \bar\alpha) \, b \, \text{vol}_{S^2} \,, \\
F_{(5)} &= 16i\,c_6^2 \biggl[ \kappa^2 X^4 \, b' \, e^{U-V} \, dw \wedge d\bar{w} +  X^{-2} \, b \, e^{U+V} \, d\mu \wedge (\partial\mathcal{G}\,dw - \bar\partial\mathcal{G}\,d\bar{w}) \biggr] \wedge \text{vol}_{\text{AdS}_3} \\
	& + \frac{2c_6^2}{3\mathcal{D}} \biggl[ \mathcal{G} X^4 \, b' \, d\mu + \frac{1}{3} \, b \, d\mathcal{G} \biggr] \wedge \text{vol}_{S^2} \wedge \text{vol}_{\tilde{S}^2} \,,
\end{split}
\end{equation}
with $\alpha=\mathcal{A}_+-\bar{\mathcal{A}}_-$ and $\beta=\mathcal{C}X^2/\sqrt{\mathcal{D}}$. The other relevant quantities are given in ~\eqref{truncationansatz6d-ffN}-\eqref{truncationansatz6d-GCD}. 

It is possible to show that the background~\eqref{uplift6dDW} is equivalent to the metric~\eqref{brane_metric_D3NS5D5NS5D5D7_nh} describing the near-horizon of the D3-NS5-D5-NS5'-D5'-D7 intersection, for a particular choice of $H_{\mathrm{D}7}$, $H_{\mathrm{D}5}$ and $H_{\mathrm{NS}5}$. First of all we extract from the equations \eqref{NS5D5D7EOMS2} an explicit form of the solution for $H_{\text{D}7}$,
\begin{equation}\label{HD7}
H_{\text{D}7} = Q_{\text{D}7} \, z \,,
\end{equation}
and impose the following contraints on the holomorphic functions featuring the dimensional reduction
\begin{equation}
\mathcal{C} = \bar{\mathcal{C}} \,,  \qquad  \mathcal{A}_+ + \bar{\mathcal{A}}_- - \bar{\mathcal{A}}_+ - \mathcal{A}_- = -\frac{3 Q_{\text{NS}5'}}{c_6\,Q_{\text{D}7}} \, \frac{\alpha - \bar{\alpha}}{\alpha + \bar{\alpha}} \,.
\end{equation}
A crucial point to relate the two Type IIB backgrounds is the following mapping between the coordinates $(z,\rho,\phi)$ of the string background \eqref{brane_metric_D3D5NS5D5NS5D7} and the parameters that define the domain wall solution, 
\begin{equation}
\label{coord6dAdS6}
z = \frac{c_6}{Q_{\mathrm{D}3}} (\alpha+\bar\alpha) \, e^U X \,,  \qquad  \rho = \frac{2 c_6^2}{Q_{\mathrm{D}3}^3} \, \mathcal{G} \, e^{3U} X^{-1} \,,  \qquad  \phi = -\frac{i}{Q_{\text{D}7}} \, \frac{\alpha-\bar\alpha}{\alpha+\bar\alpha}.
\end{equation}
From here and making use of the 6d BPS equations~\eqref{chargedDW6d}-\eqref{chargedDW6d1}, one can then express the warp factors describing the D5' and NS5' branes in~\eqref{brane_metric_D3NS5D5NS5D5D7_nh} in terms of the 6d domain wall and holomorphic quantities of the Riemann surface, as
\begin{equation}
\label{restH5HNS5}
H_{\text{D}5'} = \frac{Q_{\mathrm{D}3}^5}{18 c_6^2\,Q_{\text{NS}5'}} \, \mathcal{G}^{-1} \mathcal{D}^{-1} e^{-5U} X^3 \,,  \qquad  H_{\text{NS}5'} = \frac{Q_{\mathrm{D}3}^3\,Q_{\text{NS}5'}}{2 c_6^2} \, \mathcal{G}^{-1} e^{-3U} X = \frac{Q_{\text{NS}5'}}{\rho} \,.
\end{equation}
One can check that these expressions satisfy the equations of motion for $H_{\text{D}5'}$ and $H_{\text{NS}5'}$ written in~\eqref{NS5D5D7EOMS2}.

We have thus shown that the $\ma N=(0,4)$ AdS$_3$ background~\eqref{brane_metric_D3NS5D5NS5D5D7_nh}, describing the near-horizon limit of D3-D5-NS5 branes ending on the D5'-NS5'-D7 brane system, reproduces in a particular limit the AdS$_6$ vacuum \eqref{brane_metric_D5NS5D7_nh}, for $H_{\text{D}7}$, $H_{\text{D}5'}$ and $H_{\text{NS}5'}$ given by~\eqref{HD7} and \eqref{restH5HNS5}. Both the $\mrm{AdS}_6$ and the $\Sigma_2$ geometries arise thanks to a non-linear mixing of the $(z,\rho,\phi)$ coordinates of the brane solution \eqref{brane_metric_D3D5NS5D5NS5D7}.
The fact that the near-horizon geometry \eqref{brane_metric_D3NS5D5NS5D5D7_nh}  can be related to a 6d domain wall admitting locally AdS$_6$ in its asymptotics hints, from the supergravity point of view, at a defect interpretation. The 6d domain wall captures the effect of the D3-D5-NS5 branes intersecting  the D5'-NS5'-D7 bound state. In the domain wall coordinates, when $\mu\to0$, we approach the 6d singularity of the domain wall, which defines the locus of the defect, where the branes intersect. Instead, in the opposite limit, $\mu\to1$, a locally $\mrm{AdS}_6\times S^2 \times\Sigma_2$ vacuum with additional fluxes describing the charges of defect D3-NS5-D5 branes is recovered, for an observer far from the locus of the intersection.

\section{Conclusions}\label{conclusions}

In this paper we have constructed two new families of AdS$_3\times S^2\times {\tilde S}^2 \times S^1$ solutions to Type IIB supergravity with $(0,4)$ supersymmetries, fibrered over two intervals. These solutions are T-dual, respectively, to the class of AdS$_3\times S^2\times \text{CY}_2 \times I$ solutions to massive Type IIA supergravity constructed in \cite{Lozano:2019emq}, for the case in which $ \text{CY}_2=T^4/\mathbb{Z}_{k'}$, and to the class of AdS$_3\times S^3/\mathbb{Z}_k\times {\tilde S}^2\times I\times I'$ solutions to massless Type IIA supergravity recently constructed in \cite{Faedo:2020nol}. In the absence of D7-branes our two classes of solutions are S-dual to each other, and we have shown that, in fact, a wider class of solutions can be constructed by acting with $\text{SL}(2,\mathbb{R})$ on any of the two classes of solutions. We have discussed the M-theory realisation of the S-duality relation, identifying the AdS$_3\times S^3/\mathbb{Z}_{k}\times T^4/\mathbb{Z}_{k'}\times I$ M-theory solutions from which the two classes of solutions arise upon dimensional reduction on different directions. This class of solutions was recently constructed in \cite{Lozano:2020bxo}.

We have identified the brane intersections underlying our two classes of solutions. Both of them consist on D3-brane boxes preserving 4 Poincar\'e supersymmetries, like the ones studied in \cite{Hanany:2018hlz}. We have provided explicit supergravity solutions that describe these intersections, from which our two classes of AdS$_3$ solutions arise in the near-horizon limit. We have generalised the first intersections to include D7-branes. 
From the D3-brane boxes we have constructed explicit 2d quiver CFTs dual to our two classes of solutions. For the first class we have exploited the T-duality that relates them to the D2-D4-D6-NS5-D8 brane intersections that underlie the AdS$_3\times S^2\times \text{CY}_2 \times I$ solutions to massive Type IIA supergravity constructed in \cite{Lozano:2019emq}. The 2d quiver CFTs dual to these IIA solutions were studied in \cite{Lozano:2019zvg,Lozano:2019ywa}. We have shown that T-duality imposes a regularisation prescription for one of the interval directions that allows for a one to one mapping with the solutions in \cite{Lozano:2019emq}. Both classes of solutions then share the same 2d dual CFTs.
For the second class of solutions we have used the S-duality that relates them to the first class (in the absence of D7-branes) to identify the corresponding 2d dual CFTs. We have provided both a non-perturbative description, in terms of open D1-strings, and a perturbative one, in terms of open F1-strings.  Moreover, we have shown that the central charge is in agreement with the holographic result.

As a result of our previous analysis we have provided explicit dual CFTs to the class of AdS$_3\times S^3/\mathbb{Z}_k\times S^2\times I\times I'$ solutions to massless Type IIA supergravity constructed in \cite{Faedo:2020nol}. These solutions are related by T-duality to our second class of IIB solutions in a convenient regularisation scheme, and are thus holographically dual to the same family of 2d quiver CFTs. We have discussed the relation between these CFTs and the ones already proposed in \cite{Faedo:2020nol}, which were interpreted as surface defect CFTs within the 6d $(1,0)$ CFT dual to the AdS$_7$ solution to massless Type IIA supergravity \cite{Cvetic:2000cj}. The key result that allowed for this interpretation is the mapping found in \cite{Faedo:2020nol} between a subclass of these solutions and the 7d domain wall solution constructed in \cite{Dibitetto:2017klx}, which reproduces asymptotically locally the AdS$_7$ geometry in the UV.
We have clarified that the class of solutions whose dual CFTs have been identified in this paper are defined in terms of a function $H_{\mathrm{D}5'}$ (or $h_4$, in the notation of  \cite{Faedo:2020nol}) that must be independent of  one of the two intervals. In turn, the class of solutions considered in \cite{Faedo:2020nol} must have a non-trivial dependence on both directions. This turns out to be a crucial difference between the two families of solutions that significantly changes their CFT interpretations. In contrast with the Type IIA solutions, we have not been able to provide a surface defect interpretation for our second class of Type IIB solutions for any $H_{\mathrm{D}5'}$ function. This was to be expected given the absence of AdS$_7$ solutions in Type IIB supergravity. We have tried to link the solutions with the 6d domain wall found in \cite{Dibitetto:2018iar}, asymptotically locally AdS$_6$, which would have allowed for a surface defect interpretation within a 5d fixed point theory, but we have not succeeded. The relation between D3-brane boxes and 3d $\mathcal{N}=4$ CFTs found in \cite{Hanany:2018hlz} (see also  \cite{Chung:2016pgt,Okazaki:2019bok}) is suggestive that a defect interpretation within these CFTs could be possible.

We have seen that our first class of solutions to Type IIB allows for a surface defect interpretation when D7-branes are present. Indeed, we have linked these solutions to a 6d domain wall that reproduces asymptotically locally the AdS$_6$ solution in the general class of \cite{DHoker:2016ujz,DHoker:2017mds,DHoker:2017zwj}, that is related by T-duality to the Brandhuber-Oz solution. This has allowed us to interpret the AdS$_3$ solutions  as describing D3-D5-NS5 surface defects within the 5d CFT living in the D5-NS5-D7 brane intersection that underlies this solution. It would be interesting to provide 2d quiver CFTs that realise this explicitly, as in the constructions in \cite{Faedo:2020nol}.


\section*{Acknowledgements}

We would like to thank Antonio Amariti, Giuseppe Dibitetto, Carlos Nunez, Tadashi Okazaki and Anayeli Ramirez for very useful discussions. The authors are partially supported by the Spanish government grant PGC2018-096894-B-100 and by the Principado de Asturias through the grant FC-GRUPIN-IDI/2018/000174.

\appendix

\section{Summary of the 2d CFTs dual to the AdS$_3\times$ S$^2\times \text{CY}_2 \times I_z$ solutions 
}
\label{summary2dCFT}

In this Appendix we briefly summarise the 2d CFTs dual to the AdS$_3\times S^2\times \text{CY}_2 \times I_z$ solutions in \cite{Lozano:2019emq}.
A generic background in the class of solutions constructed in \cite{Lozano:2019emq} is defined by three different functions, $h_4,h_8, u$
\footnote{There is actually a more general class depending on an extra $H_2$ function in the $\text{CY}_2$, but we are not concerned with this wider class of solutions in this paper.}. The functions $h_8$ and $u$ depend only on $z$, while $h_4$ depends on both $z$ and the coordinates of the $\text{CY}_2$. In the particular case in which $h_4$ is taken to be independent of the $\text{CY}_2$ this function becomes also a linear function of $z$. The 2d CFTs dual to this particular class of solutions were identified in \cite{Lozano:2019jza,Lozano:2019zvg,Lozano:2019ywa}.

The 2d quiver CFTs constructed in these references are in one to one correspondence with the  quantised charges associated to the solutions specified above. These were computed in \cite{Lozano:2019zvg} (see equation (2.16) therein). Specifying to the case in which $\text{CY}_2=T^4/\mathbb{Z}_{k'}$, our  interest in this paper, we find\footnote{We take also $u$ to be constant for simplicity. See \cite{Dibitetto:2020bsh} for the brane interpretation in M-theory when the function $u$ is not a constant.}
\begin{align}
N_6 &= \frac{1}{2\pi}\int_{S^2}{\hat F}_{(2)}=h_8=k\,, \label{D6charge1}\\
N_4 &= \frac{1}{(2\pi)^3}\int_{T^4/\mathbb{Z}_{k'}}{\hat F}_{(4)}=\frac{2\pi}{k'}h_4'\,, \label{D4charge1} \\
N_2 &= \frac{1}{(2\pi)^5}\int_{T^4/\mathbb{Z}_{k'}\times S^2} {\hat F}_{(6)}=\frac{1}{k'}\Bigl(h_4-(z-2\pi j)h_4'\Bigr) \,,\label{D2charge1} \\
N_{\mathrm{NS}5} &= \frac{1}{(2\pi)^2}\int_{I_z\times S^2} H_{(3)}=1 \label{NS5charge1}\, ,
\end{align}
where primes denote derivatives with respect to $z$ and we have used $h_8= k$~\footnote{This is the number of KK-monopoles in the original M-theory set-up.}.
These charges were computed in the interval $2\pi j \leq z \leq 2\pi (j+1)$. As explained in \cite{Lozano:2019zvg}, the $z$ direction must be divided in intervals of length $2\pi$, such that a large gauge transformation of parameter $j$ enforces that the condition $\frac{1}{4\pi^2}\oint_{S^2}B_{(2)}\in [0,1)$ holds. General quiver CFTs can then be constructed taking $h_4(z)$ piecewise continuous across these intervals, as 
 \begin{equation} \label{profileh4sp2}
\frac{1}{k'}\,h_4(z) =
\left\{
\begin{array}{ccl}
  \frac{\beta_0 }{2\pi}z    && 0\leq z\leq 2\pi \\
  \alpha_j+\frac{\beta_j}{2\pi}(z-2\pi j)    && 2\pi j \leq z \leq 2\pi (j+1) \,,  \quad  j=1,\dots P-1 \\
  \alpha_P-\frac{\alpha_P}{2\pi}(z-2\pi P)    && 2\pi P\leq z \leq 2\pi(P+1) \,,
\end{array}
\right.
\end{equation}
where 
\begin{equation}
\alpha_j=\sum_{i=0}^{j-1}\beta_i\, .
\end{equation}
The choice of constants is imposed by continuity of the metric and dilaton. Interestingly, this choice implies discontinuities in the RR-sector, that are interpreted as generated by D4-brane sources in the background. A crucial ingredient of the quiver constructions in \cite{Lozano:2019zvg} is that they describe well-defined, finite, 2d CFTs. This requires a global definition of the $I_z$-direction. In the massless case this is achieved requiring that $h_4(z)$ vanishes at both ends of the $z$-interval, that are taken at $z=0,2\pi (P+1)$. This introduces extra D4-branes in the configuration. For $h_4$ given by (\ref{profileh4sp2}) we have
\begin{align}
N_6^{(j)} &= \frac{1}{2\pi}\int_{S^2}{\hat F}_{(2)}=h_8=k\,, \label{D6charge2}\\
N_4^{(j)} &= \frac{1}{(2\pi)^3}\int_{T^4/\mathbb{Z}_{k'}}{\hat F}_{(4)}= \beta_j\,, \label{D4charge2} \\
N_2^{(j)} &= \frac{1}{(2\pi)^5}\int_{T^4/\mathbb{Z}_{k'}\times S^2} {\hat F}_{(6)}= \alpha_j \,,\label{D2charge2} \\
N_{\mathrm{NS}5}^{(j)} &= \frac{1}{(2\pi)^2}\int_{I_z\times S^2} H_{(3)}=1 \label{NS5charge2}\, .
\end{align}
These equations imply that $\beta_j$ and $\alpha_j$ must be integer numbers.

The backgrounds defined by the functions $h_4,h_8$ specified above are dual to the CFTs describing the low-energy dynamics of the two-dimensional quantum field theories encoded by the  quivers in Figure \ref{figurageneral}.
\begin{figure}[h!]
\centering
\includegraphics[width=9cm]{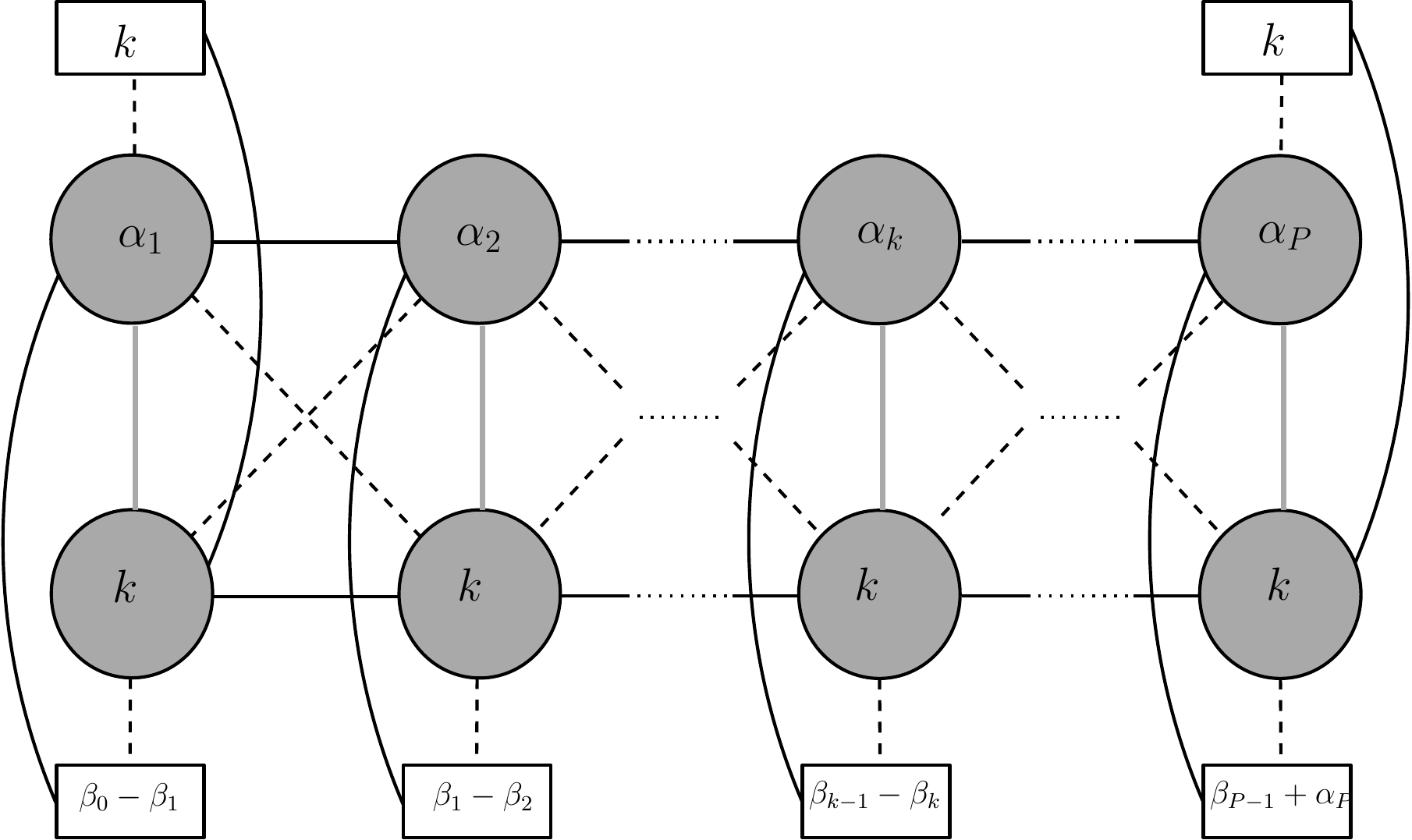}%
\caption{A generic quiver field theory whose IR is dual to the holographic background defined by $h_8=k$ and $h_4$ as in \eqref{profileh4sp2}. The solid black lines represent $(4,4)$ hypermultiplets, the grey lines represent $(0,4)$ hypermultiplets and the dashed lines represent $(0,2)$ Fermi multiplets, all of them in the bifundamental representations of the gauge groups. ${\cal N}=(4,4)$ vector multiplets are the degrees of freedom  in each gauge node.}
\label{figurageneral}
\end{figure}
In these quivers the gauge groups are associated to stacks of D2 and D6 branes (the latter wrapped on $T^4/\mathbb{Z}_{k'}$) stretched between NS5-branes. Figure \ref{brane-set-up1} illustrates the 
corresponding Hanany-Witten brane set-ups.
 \begin{figure}
\centering
\includegraphics[scale=0.7]{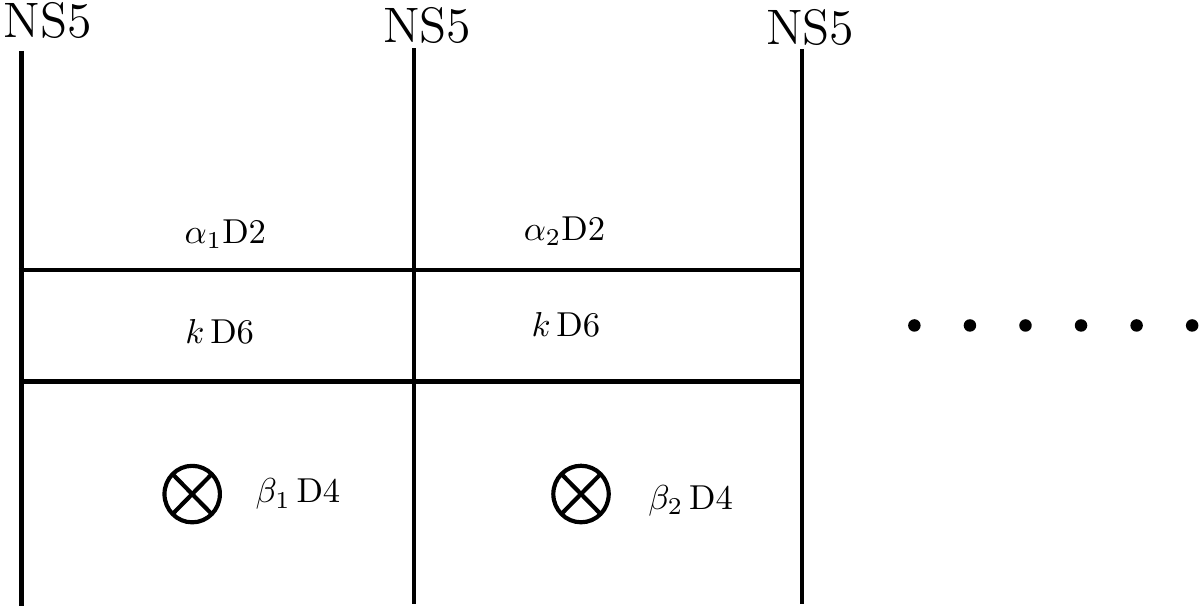}
\caption{Generic brane set-up associated to the AdS$_3$ solutions in \cite{Lozano:2019emq} with vanishing Romans mass.}
\label{brane-set-up1}
\end{figure} 
The quivers are \textit{planar}, in that they consist on two long linear quivers, built out of the gauge groups associated to the D2 and D6 branes, coupled by matter multiplets. Each linear quiver consists on $(4,4)$ gauge groups connected horizontally by $(4,4)$ bifundamental hypermultiplets. They couple to each other through $(0,4)$ hypermultiplets (vertically) and $(0,2)$ Fermi multiplets (in the diagonals). These render the final \textit{planar} quiver CFTs $(0,4)$ supersymmetric. 
Note that, since the 2d theory is chiral, one needs to be careful with gauge anomaly cancellation. This is ensured thanks to the D4-branes present in each interval, which precisely provide the flavour groups at each node required by anomaly cancelation (see below), 
\begin{equation} \label{efes}
F_j= \beta_{j-1}-\beta_{j}.
\end{equation}
These flavour groups couple through $(0,2)$ Fermi multiplets with the corresponding gauge nodes. 

Finally, we recall briefly how the cancelation of gauge anomalies works in $(0,4)$ 2d CFTs. For a given $\mathrm{U}(\alpha_j)$  ($\mathrm{U}(k)$) gauge group the contributions to the gauge anomaly come from the $(0,4)$ hypermultiplet and $(0,2)$ Fermi multiplets in the bifundamental representations that connect it to the $\mathrm{U}(k)$ ($\mathrm{U}(\alpha_{j-1})$, $\mathrm{U}(\alpha_j)$, $\mathrm{U}(\alpha_{j+1})$) gauge groups. The $(0,4)$ hypermultiplets are composed of two $(0,2)$ chiral multiplets, which contribute to the anomaly a factor of 1. In turn, the $(0,2)$ Fermi multiplets contribute a factor of $-\frac12$. Putting these together, we have that for a $\mathrm{U}(\alpha_j)$ gauge group the gauge anomaly is cancelled without the addition of any flavour groups\footnote{Other than in the original and final gauge groups.}, while for the $\mathrm{U}(k)$ gauge groups it is necessary to add $2\alpha_j-\alpha_{j+1}-\alpha_{j-1}=\beta_{j-1}-\beta_{j}$ Fermi multiplets, which is precisely the number of flavour groups given by equation \eqref{efes}.

\subsection{Central charge}

The most stringent check in \cite{Lozano:2019jza,Lozano:2019zvg,Lozano:2019ywa} for the validity of the proposed quivers was the matching between the field theory and holographic central charges. The U(1)$_R$ current two-point function can be identified when flowing to the far IR with the right moving central charge of the ${\cal N}=(0,4)$ conformal field theory. Following \cite{Putrov:2015jpa} it was found that for a generic quiver with $n_{hyp}$ hypermultiplets and $n_{vec}$ vector multiplets, the central charge (the anomaly of the U(1)$_R$ current) is
\begin{equation} \label{centralcft}
c_{CFT}=6 (n_{hyp}- n_{vec}) \,.
\end{equation}
In \cite{Lozano:2019zvg,Lozano:2019ywa} a variety of examples of long linear quivers for which the rank of each of the nodes is a large number were discussed. In each of these qualitatively different examples, it was found that the field theoretical central charge of equation~\eqref{centralcft} coincides (in the limit of long quivers with large ranks) with the holographic central charge, computed as
\begin{equation} \label{chol}
c_{hol}= \frac{3\pi}{2\,G_\text{N}}\,\text{Vol}_{T^4/\mathbb{Z}_{k'}} \int_0^{2\pi (P+1)} h_4 h_8 \, d z= \frac{3}{\pi k'}  \int_0^{2\pi (P+1)} {h}_4 h_8 \, d z \,.
\end{equation}
In the massless case $h_8$ is constant. Substituting the expression for $h_4$ given by~\eqref{profileh4sp2} we obtain 
\begin{equation}
c_{hol}=6 k \sum_{j=1}^{P}\alpha_j \,.
\end{equation}

\section{6d gauged supergravity from Type IIB on $S^2\times \Sigma_2$}
\label{6dsugra}

In this Appendix we briefly discuss the consistent truncation of Type IIB string theory around the family of $\text{AdS}_6 \times S^2\times \Sigma_2$ vacua associated to $(p,q)$ 5-branes and 7-branes first classified in~\cite{DHoker:2016ujz,DHoker:2017mds,DHoker:2017zwj}. This consistent truncation gives rise to the $F(4)$ minimal gauged supergravity introduced in \cite{Romans:1985tw}. This theory preserves 16 real supercharges, giving rise to $\mathcal{N}=(1,1)$ supersymmetry in six dimensions, and is commonly referred to as Romans supergravity. The field content is given by the 6d gravitational field, a real scalar $X$, a 2-form gauge potential $\mathcal{B}_{(2)}$, three $\text{SU(2)}$ vectors $\mathcal{A}^i$ and one Abelian vector $\mathcal{A}^0$. Here we will restrict to the case of vanishing vector fields. 

The truncation from IIB supergravity to Romans supergravity was worked out in~\cite{Hong:2018amk}. In what follows we summarise the main results of this article assuming vanishing 6d vectors. The ansatz for the metric, here given in the Einstein frame, is characterised by an internal manifold locally realised as a fibration of a 2-sphere over a Riemann surface\footnote{We use the notation $\tilde S^2$ to be compatible with section \ref{defects}.} $\Sigma_2$, parametrised by the complex coordinate $w$ \cite{Hong:2018amk},
\begin{equation}
\label{truncationansatz6d}
ds_{10}^2 = f_6^2 \, ds_6^2 + f_2^2 \, ds_{\tilde{S}^2}^2 + 4N^2 \, dw d\bar{w} \,,
\end{equation}
where
\begin{equation}
\label{truncationansatz6d-ffN}
f_6^2 = \frac{c_6^2}{N^2} \, \kappa^2 \sqrt{\mathcal{D}} \,,  \qquad  f_2^2 = \frac{c_6^2}{9N^2} \frac{\kappa^2 X^2}{\sqrt{\mathcal{D}}} \,,  \qquad  N^4 = \frac{c_6^2}{6} \frac{\kappa^4 X^2 \sqrt{\mathcal{D}}}{\mathcal{G}} \,,
\end{equation}
with $c_6$ a non-vanishing constant.
The reduction ansatz is completely fixed by a pair of holomorphic functions $\mathcal{A}_\pm(w)$, from which we define a holomorphic function $\mathcal{B}(w)$ whose derivative
is given by\footnote{We use the notation $\partial=\partial_w$ and $\bar \partial=\partial_{\bar w}$.}
\begin{equation}
\partial\mathcal{B} = \mathcal{A}_+ \partial\mathcal{A}_- - \mathcal{A}_- \partial\mathcal{A}_+ \,.
\end{equation}
From $\mathcal{A}_\pm$ and $\mathcal{B}$ we define
\begin{equation}
\kappa_\pm = \partial \mathcal{A}_\pm \,,  \qquad  \kappa^2 = -|\kappa_+|^2 + |\kappa_-|^2 \,,
\end{equation}
along with
\begin{equation}
\label{truncationansatz6d-GCD}
\mathcal{G} = |\mathcal{A}_+|^2 - |\mathcal{A}_-|^2 + \mathcal{B} + \bar{\mathcal{B}} \,,  \qquad  \mathcal{C} = \frac{\kappa_+ \bar{\partial}\mathcal{G} + \bar{\kappa}_- \partial\mathcal{G}}{\kappa^2} \,,  \qquad  \mathcal{D} = X^4 + \frac23 \frac{|\partial\mathcal{G}|^2}{\kappa^2 \mathcal{G}} \,.
\end{equation}
The truncation ansatz for the gauge potential is expressed in complex notation in terms of a scalar $B$ and a complex 2-form $\tilde C_2$ as follows \cite{Hong:2018amk}
\begin{equation}
\begin{split}
B & = \frac{(\mathcal{A}_+ - \bar{\mathcal{A}}_-) - \mathcal{C} X^2/\sqrt{\mathcal{D}}}{(\bar{\mathcal{A}}_+ - \mathcal{A}_-) + \bar{\mathcal{C}} X^2/\sqrt{\mathcal{D}}} \,, \\
\tilde C_2 & = i\,\frac{2c_6}{9} \biggl[ \frac{\mathcal{C}}{\mathcal{D}} - 3(\mathcal{A}_+ + \bar{\mathcal{A}}_-) \biggr] \, \text{vol}_{\tilde{S}^2} - 2\, c_6 (\mathcal{A}_+ - \bar{\mathcal{A}}_-) \, \mathcal{B}_{(2)} \,.
\end{split}
\end{equation}
We can rewrite the aforementioned truncation ansatz in the real notation and in the string frame in order to apply it to the purposes of this paper, namely \cite{Grana:2001xn}
\begin{equation}
\begin{split}
\tau & = C_{(0)} + i e^{-\Phi}  \qquad  \text{with}  \qquad  \tau = i \, \frac{1 - B}{1 + B} \,, \\
\tilde C_2 & = B_{(2)} + i C_{(2)} \,.
\end{split}
\end{equation}
The metric in the string frame is given, as usual, by $ds_{10, s}^2  = e^{\Phi/2} ds_{10,\text{E}}^2$. In real notation the 10d fluxes are decomposed as\footnote{We added a factor of 4 to the 5-form flux in order to be consistent with our conventions.}
\begin{equation}
\label{10dfluxesto6d}
\begin{split}
H_{(3)} = & \ dB_{(2)} \,,  \qquad  F_{(1)} = dC_{(0)} \,,  \qquad  F_{(3)} = dC_{(2)} - H_{(3)} C_{(0)} \,, \\
F_{(5)} = & \ 4c_6^2 \biggl[ \frac{i\,\kappa^2 X^4}{2} \star_6\!\mathcal{F}_{(3)} \wedge dw \wedge d\bar{w} + \frac{1}{2 X^2} \star_6\!\mathcal{B}_{(2)} \wedge \star_2 d\mathcal{G} \\
	  & + \frac{\mathcal{G} X^4}{6\mathcal{D}} \, d\mathcal{B}_{(2)} \wedge \text{vol}_{\tilde{S}^2} + \frac{m}{18 \mathcal{D}} \, \mathcal{B}_{(2)} \wedge d\mathcal{G} \wedge \text{vol}_{\tilde{S}^2} \biggr] \,,
\end{split}
\end{equation}
where $\mathcal{F}_{(3)}=d\mathcal{B}_{(2)}$.
The 6d theory resulting from this truncation preserves 16 real supercharges and its gauging is realised by the $\mathrm{SU}(2)_R$ R-symmetry group and by a mass deformation of the 2-form $\ma B_{(2)}$. From a purely six-dimensional point of view, the theory can thus be formulated in terms of two independent gauge parameters, namely $g$ and $m$, taking into account the two different gauge deformations. Here we follow the particular choice of \cite{Hong:2018amk}, where the St\"uckelberg parameter $m$ is fixed to one and $g=\frac{3}{\sqrt 2}$ in order to have an $\mrm{AdS}_6$ vacuum of unit radius.
In particular, the truncation ansatz~\eqref{truncationansatz6d} produces a scalar potential in six dimensions defined by the superpotential
\begin{equation}
\label{6dsuperpotential}
f(g,X) = \frac18 \left(X^{-3} + 3\,X\right) \,.
\end{equation}
The 6d Lagrangian has the form
\begin{equation}
\begin{split}
\label{6dlagrangian}
\mathcal{L}_6 & = R_{\,6} - 4\,X^{-2} \star_6\!dX \wedge dX - \frac12 \, X^4 \star_6\!\mathcal{F}_{(3)} \wedge \mathcal{F}_{(3)} - V \\
 & - X^{-2} \star_6\!\mathcal{B}_{(2)} \wedge \mathcal{B}_{(2)} - \frac13  \, \mathcal{B}_{(2)} \wedge \mathcal{B}_{(2)} \wedge \mathcal{B}_{(2)} \,,
\end{split}
\end{equation}
where the scalar potential is given by
\begin{equation}
V = 16\,X^2 \left(D_X f\right)^2 - 80\,f^2 \,.
\end{equation}
The 6d theory defined by~\eqref{6dlagrangian} has a supersymmetric AdS$_6$ vacuum at $X=1$ and vanishing gauge potentials. This vacuum preserves 16 supercharges and it is associated to the whole family of $\text{AdS}_6 \times S^2 \times \Sigma_2$ solutions introduced in \cite{DHoker:2016ujz,DHoker:2017mds,DHoker:2017zwj}.


\begin{thebibliography}{99}
\bibliographystyle{unsrt}

\bibitem{Couzens:2017way}
  C.~Couzens, C.~Lawrie, D.~Martelli, S.~Schafer-Nameki and J.~M.~Wong,
  ``F-theory and AdS$_{3}$/CFT$_{2}$,''
  JHEP \textbf{08} (2017), 043
  [arXiv:1705.04679 [hep-th]].

\bibitem{Macpherson:2018mif}
  N.~T.~Macpherson,
  ``Type II solutions on AdS$_{3} \times$ S$^{3} \times$ S$^{3}$ with large superconformal symmetry,''
  JHEP \textbf{05} (2019), 089
  [arXiv:1812.10172 [hep-th]].

\bibitem{Lozano:2019emq}
  Y.~Lozano, N.~T.~Macpherson, C.~Nunez and A.~Ramirez,
  ``AdS$_3$ solutions in Massive IIA with small $\mathcal{N}=(4,0)$ supersymmetry'',
  JHEP \textbf{01}, 129 (2020)
  [arXiv:1908.09851 [hep-th]].
 
\bibitem{Lozano:2020bxo}
  Y.~Lozano, C.~Nunez, A.~Ramirez and S.~Speziali,
  ``$M$-strings and AdS$_3$ solutions to M-theory with small $\mathcal{N}=(0,4)$ supersymmetry,''
  JHEP \textbf{08}, 118 (2020)
  [arXiv:2005.06561 [hep-th]].
 
\bibitem{Faedo:2020nol}
  F.~Faedo, Y.~Lozano and N.~Petri,
  ``Searching for surface defect CFTs within AdS$_3$,''
  JHEP \textbf{11}, 052 (2020)
  [arXiv:2007.16167 [hep-th]].

\bibitem{Lozano:2019jza} 
  Y.~Lozano, N.~T.~Macpherson, C.~Nunez and A.~Ramirez,
  ``1/4 BPS solutions and the AdS$_3$/CFT$_2$ correspondence,''
  Phys.\ Rev.\ D {\bf 101}, no. 2, 026014 (2020)
  [arXiv:1909.09636 [hep-th]].

\bibitem{Lozano:2019zvg} 
  Y.~Lozano, N.~T.~Macpherson, C.~Nunez and A.~Ramirez,
  ``Two dimensional ${\cal N}=(0,4)$ quivers dual to AdS$_3$ solutions in massive IIA,''
  JHEP {\bf 2001}, 140 (2020)
  [arXiv:1909.10510 [hep-th]].
  
\bibitem{Lozano:2019ywa} 
  Y.~Lozano, N.~T.~Macpherson, C.~Nunez and A.~Ramirez,
  ``AdS$_3$ solutions in massive IIA, defect CFTs and T-duality,''
  JHEP {\bf 1912}, 013 (2019)
  [arXiv:1909.11669 [hep-th]].
  
\bibitem{Boonstra:1998yu}
H.~J.~Boonstra, B.~Peeters and K.~Skenderis,
``Brane intersections, anti-de Sitter space-times and dual superconformal theories,''
Nucl. Phys. B \textbf{533}, 127-162 (1998)
[arXiv:hep-th/9803231 [hep-th]].
  
\bibitem{Maldacena:1997de}
  J.~M.~Maldacena, A.~Strominger and E.~Witten,
  ``Black hole entropy in M theory,''
  JHEP \textbf{12} (1997), 002
  [arXiv:hep-th/9711053 [hep-th]].

\bibitem{Vafa:1997gr}
  C.~Vafa,
  ``Black holes and Calabi-Yau threefolds,''
  Adv. Theor. Math. Phys. \textbf{2} (1998), 207-218
  [arXiv:hep-th/9711067 [hep-th]].

\bibitem{Minasian:1999qn}
  R.~Minasian, G.~W.~Moore and D.~Tsimpis,
  ``Calabi-Yau black holes and (0,4) sigma models,''
  Commun. Math. Phys. \textbf{209} (2000), 325-352
  [arXiv:hep-th/9904217 [hep-th]].

\bibitem{Castro:2008ne}
  A.~Castro, J.~L.~Davis, P.~Kraus and F.~Larsen,
  ``String Theory Effects on Five-Dimensional Black Hole Physics,''
  Int. J. Mod. Phys. A \textbf{23} (2008), 613-691
  [arXiv:0801.1863 [hep-th]].

\bibitem{Haghighat:2015ega}
  B.~Haghighat, S.~Murthy, C.~Vafa and S.~Vandoren,
  ``F-Theory, Spinning Black Holes and Multi-string Branches,''
  JHEP \textbf{01} (2016), 009
  [arXiv:1509.00455 [hep-th]].

\bibitem{Couzens:2019wls}
  C.~Couzens, H.~het Lam, K.~Mayer and S.~Vandoren,
  ``Black Holes and (0,4) SCFTs from Type IIB on K3,''
  JHEP \textbf{08} (2019), 043
  [arXiv:1904.05361 [hep-th]].

\bibitem{Couzens:2020aat}
  C.~Couzens, H.~het Lam, K.~Mayer and S.~Vandoren,
  ``Anomalies of (0,4) SCFTs from F-theory,''
  JHEP \textbf{08} (2020), 060
  [arXiv:2006.07380 [hep-th]].

\bibitem{Haghighat:2013tka}
  B.~Haghighat, C.~Kozcaz, G.~Lockhart and C.~Vafa,
  ``Orbifolds of M-strings,''
  Phys. Rev. D \textbf{89} (2014) no.4, 046003
  [arXiv:1310.1185 [hep-th]].

\bibitem{Gadde:2015tra}
  A.~Gadde, B.~Haghighat, J.~Kim, S.~Kim, G.~Lockhart and C.~Vafa,
  ``6d String Chains,''
  JHEP \textbf{02} (2018), 143
  [arXiv:1504.04614 [hep-th]].

\bibitem{Karch:2000gx} 
  A.~Karch and L.~Randall,
  ``Open and closed string interpretation of SUSY CFT's on branes with boundaries,''
  JHEP {\bf 0106}, 063 (2001)
  [hep-th/0105132].

\bibitem{DeWolfe:2001pq} 
  O.~DeWolfe, D.~Z.~Freedman and H.~Ooguri,
  ``Holography and defect conformal field theories,''
  Phys.\ Rev.\ D {\bf 66}, 025009 (2002)
  [hep-th/0111135].

\bibitem{Bachas:2001vj} 
  C.~Bachas, J.~de Boer, R.~Dijkgraaf and H.~Ooguri,
  ``Permeable conformal walls and holography,''
  JHEP {\bf 0206}, 027 (2002)
  [hep-th/0111210].

\bibitem{Erdmenger:2002ex} 
  J.~Erdmenger, Z.~Guralnik and I.~Kirsch,
  ``Four-dimensional superconformal theories with interacting boundaries or defects,''
  Phys.\ Rev.\ D {\bf 66}, 025020 (2002)
  [hep-th/0203020].

\bibitem{Clark:2004sb} 
  A.~B.~Clark, D.~Z.~Freedman, A.~Karch and M.~Schnabl,
  ``Dual of the Janus solution: An interface conformal field theory,''
  Phys.\ Rev.\ D {\bf 71}, 066003 (2005)
  [hep-th/0407073].

\bibitem{Lunin:2007ab} 
  O.~Lunin,
  ``1/2-BPS states in M theory and defects in the dual CFTs,''
  JHEP {\bf 0710}, 014 (2007)
  [arXiv:0704.3442 [hep-th]].

\bibitem{DHoker:2007zhm} 
  E.~D'Hoker, J.~Estes and M.~Gutperle,
  ``Exact half-BPS Type IIB interface solutions. I. Local solution and supersymmetric Janus,''
  JHEP {\bf 0706}, 021 (2007)
  [arXiv:0705.0022 [hep-th]].

\bibitem{Gaiotto:2008sd} 
  D.~Gaiotto and E.~Witten,
  ``Janus Configurations, Chern-Simons Couplings, And The theta-Angle in N=4 Super Yang-Mills Theory,''
  JHEP {\bf 1006}, 097 (2010)
  [arXiv:0804.2907 [hep-th]].

\bibitem{Jensen:2013lxa} 
  K.~Jensen and A.~O'Bannon,
  ``Holography, Entanglement Entropy, and Conformal Field Theories with Boundaries or Defects,''
  Phys.\ Rev.\ D {\bf 88}, no. 10, 106006 (2013)
  [arXiv:1309.4523 [hep-th]].

\bibitem{Dibitetto:2017tve} 
  G.~Dibitetto and N.~Petri,
  ``BPS objects in D = 7 supergravity and their M-theory origin,''
  JHEP {\bf 1712}, 041 (2017)
  [arXiv:1707.06152 [hep-th]].
  
\bibitem{Dibitetto:2017klx} 
  G.~Dibitetto and N.~Petri,
  ``6d surface defects from massive type IIA,''
  JHEP {\bf 1801}, 039 (2018)
  [arXiv:1707.06154 [hep-th]].

\bibitem{Dibitetto:2018iar} 
  G.~Dibitetto and N.~Petri,
  ``Surface defects in the D4 $-$ D8 brane system,''
  JHEP {\bf 1901}, 193 (2019)
  [arXiv:1807.07768 [hep-th]].
  
\bibitem{Chen:2019qib}
  K.~Chen and M.~Gutperle,
  ``Holographic line defects in F(4) gauged supergravity,''
  Phys. Rev. D \textbf{100}, no.12, 126015 (2019)
  [arXiv:1909.11127 [hep-th]].
  
\bibitem{Chen:2020mtv}
  K.~Chen, M.~Gutperle and M.~Vicino,
  ``Holographic Line Defects in $D=4$, $N=2$ Gauged Supergravity,''
  Phys. Rev. D \textbf{102}, no.2, 026025 (2020)
  [arXiv:2005.03046 [hep-th]].

\bibitem{Seiberg:1996bd}
  N.~Seiberg,
  ``Five-dimensional SUSY field theories, nontrivial fixed points and string dynamics,''
  Phys. Lett. B \textbf{388} (1996), 753-760
  [arXiv:hep-th/9608111 [hep-th]].

\bibitem{Brandhuber:1999np} 
  A.~Brandhuber and Y.~Oz,
  ``The D-4 - D-8 brane system and five-dimensional fixed points,''
  Phys.\ Lett.\ B {\bf 460}, 307 (1999)
  [hep-th/9905148].
 
\bibitem{Cvetic:2000cj} 
  M.~Cvetic, H.~Lu, C.~N.~Pope and J.~F.~Vazquez-Poritz,
  ``AdS in warped space-times,''
  Phys.\ Rev.\ D {\bf 62}, 122003 (2000)
  [hep-th/0005246].
  
\bibitem{Haghighat:2013gba}
  B.~Haghighat, A.~Iqbal, C.~Koz\c{c}az, G.~Lockhart and C.~Vafa,
  ``M-Strings,''
  Commun. Math. Phys. \textbf{334} (2015) no.2, 779-842
  [arXiv:1305.6322 [hep-th]].
  
\bibitem{Hanany:2018hlz}
A.~Hanany and T.~Okazaki,
``(0,4) brane box models,''
JHEP \textbf{03} (2019), 027
[arXiv:1811.09117 [hep-th]].

\bibitem{Hanany:1997tb}
A.~Hanany and A.~Zaffaroni,
``On the realization of chiral four-dimensional gauge theories using branes,''
JHEP \textbf{05} (1998), 001
[arXiv:hep-th/9801134 [hep-th]].

\bibitem{Hanany:1998it}
A.~Hanany and A.~M.~Uranga,
``Brane boxes and branes on singularities,''
JHEP \textbf{05} (1998), 013
[arXiv:hep-th/9805139 [hep-th]].

\bibitem{Bergman:2012kr}
  O.~Bergman and D.~Rodriguez-Gomez,
  ``5d quivers and their AdS(6) duals,''
  JHEP \textbf{07} (2012), 171
  [arXiv:1206.3503 [hep-th]].
 
\bibitem{Hong:2018amk}
  J.~Hong, J.~T.~Liu and D.~R.~Mayerson,
  ``Gauged Six-Dimensional Supergravity from Warped IIB Reductions'',
  JHEP \textbf{09}, 140 (2018)
  [arXiv:1808.04301 [hep-th]].
  
\bibitem{Jeong:2013jfc}
  J.~Jeong, O.~Kelekci and E.~O Colgain,
  ``An alternative IIB embedding of F(4) gauged supergravity,''
  JHEP \textbf{05} (2013), 079
  [arXiv:1302.2105 [hep-th]].
 
\bibitem{Klebanov:2007ws} 
  I.~R.~Klebanov, D.~Kutasov and A.~Murugan,
  ``Entanglement as a probe of confinement,''
  Nucl.\ Phys.\ B {\bf 796}, 274 (2008)
  [arXiv:0709.2140 [hep-th]].
  
\bibitem{Macpherson:2014eza} 
  N.~T.~Macpherson, C.~Nunez, L.~A.~Pando Zayas, V.~G.~J.~Rodgers and C.~A.~Whiting,
  ``Type IIB supergravity solutions with AdS$_{5}$ from Abelian and non-Abelian T dualities,''
  JHEP {\bf 1502}, 040 (2015)
  [arXiv:1410.2650 [hep-th]].

\bibitem{Bea:2015fja} 
  Y.~Bea, J.~D.~Edelstein, G.~Itsios, K.~S.~Kooner, C.~Nunez, D.~Schofield and J.~A.~Sierra-Garcia,
  ``Compactifications of the Klebanov-Witten CFT and new AdS$_{3}$ backgrounds,''
  JHEP {\bf 1505}, 062 (2015)
  [arXiv:1503.07527 [hep-th]].
   
\bibitem{Putrov:2015jpa}
  P.~Putrov, J.~Song and W.~Yan,
  ``(0,4) dualities,''
  JHEP \textbf{03} (2016), 185
  [arXiv:1505.07110 [hep-th]].

\bibitem{Lozano:2020txg}
Y.~Lozano, C.~Nunez, A.~Ramirez and S.~Speziali,
``New AdS$_2$ backgrounds and ${\cal N}=4$ Conformal Quantum Mechanics,''
[arXiv:2011.00005 [hep-th]].

\bibitem{Aharony:1997ju}
  O.~Aharony and A.~Hanany,
  ``Branes, superpotentials and superconformal fixed points,''
  Nucl. Phys. B \textbf{504} (1997), 239-271
  [arXiv:hep-th/9704170 [hep-th]].

\bibitem{Kim:2007hv}
  H.~Kim, K.~K.~Kim and N.~Kim,
  ``1/4-BPS M-theory bubbles with SO(3) x SO(4) symmetry,''
  JHEP \textbf{08} (2007), 050
  [arXiv:0706.2042 [hep-th]].
  
\bibitem{Apruzzi:2013yva}
  F.~Apruzzi, M.~Fazzi, D.~Rosa and A.~Tomasiello,
  ``All AdS$_7$ solutions of type II supergravity,''
  JHEP \textbf{04} (2014), 064
  [arXiv:1309.2949 [hep-th]].

\bibitem{DHoker:2016ujz}
  E.~D'Hoker, M.~Gutperle, A.~Karch and C.~F.~Uhlemann,
  ``Warped $AdS_6\times S^2$ in Type IIB supergravity I: Local solutions,''
  JHEP \textbf{08}, 046 (2016)
  [arXiv:1606.01254 [hep-th]].

\bibitem{DHoker:2017mds}
  E.~D'Hoker, M.~Gutperle and C.~F.~Uhlemann,
  ``Warped $AdS_6\times S^2$ in Type IIB supergravity II: Global solutions and five-brane webs,''
  JHEP \textbf{05}, 131 (2017)
  [arXiv:1703.08186 [hep-th]].

\bibitem{DHoker:2017zwj}
  E.~D'Hoker, M.~Gutperle and C.~F.~Uhlemann,
  ``Warped $AdS_6\times S^2$ in Type IIB supergravity III: Global solutions with seven-branes,''
  JHEP \textbf{11}, 200 (2017)
  [arXiv:1706.00433 [hep-th]].

\bibitem{Lozano:2012au}
  Y.~Lozano, E.~\'O Colg\'ain, D.~Rodr\'\i{}guez-G\'omez and K.~Sfetsos,
  ``Supersymmetric $AdS_6$ via T Duality,''
  Phys. Rev. Lett. \textbf{110} (2013) no.23, 231601
  [arXiv:1212.1043 [hep-th]].
  
\bibitem{Lozano:2018pcp}
  Y.~Lozano, N.~T.~Macpherson and J.~Montero,
  ``AdS$_{6}$ T-duals and type IIB AdS$_{6} \times$ S$^{2}$ geometries with 7-branes,''
  JHEP \textbf{01}, 116 (2019)
  [arXiv:1810.08093 [hep-th]].

\bibitem{Dibitetto:2020bsh}
  G.~Dibitetto and N.~Petri,
  ``AdS$_3$ from M-branes at conical singularities,''
  [arXiv:2010.12323 [hep-th]].

\bibitem{Romans:1985tw} 
  L.~J.~Romans,
  ``The F(4) Gauged Supergravity in Six-dimensions,''
  Nucl.\ Phys.\ B {\bf 269}, 691 (1986).

\bibitem{Grana:2001xn}
  M.~Grana and J.~Polchinski,
  ``Gauge / gravity duals with holomorphic dilaton,''
  Phys. Rev. D \textbf{65}, 126005 (2002)
  [arXiv:hep-th/0106014 [hep-th]].

\bibitem{Chung:2016pgt}
H.~J.~Chung and T.~Okazaki,
``(2,2) and (0,4) supersymmetric boundary conditions in 3d $\mathcal{N}$ = 4 theories and type IIB branes,''
Phys. Rev. D \textbf{96} (2017) no.8, 086005
[arXiv:1608.05363 [hep-th]].

\bibitem{Okazaki:2019bok}
T.~Okazaki,
``Abelian dualities of $\mathcal{N}=(0,4)$ boundary conditions,''
JHEP \textbf{08} (2019), 170
[arXiv:1905.07425 [hep-th]].



\end{thebibliography}
\end{document}